\documentclass[journal]{IEEEtran}
\usepackage{hyperref}
\usepackage{graphicx}
\usepackage{CJKutf8}
\usepackage{amsmath}
\usepackage{amssymb}
\usepackage{tabularx}
\usepackage{multirow}

\ifCLASSINFOpdf
\else
\fi

\newcommand{\Fig}[1]{Figure~\ref{fig:#1}} 
\newcommand{\Table}[1]{Table~\ref{tab:#1}} 
\newcommand{\Eq}[1]{Eq.~(\ref{eq:#1})} 
\newcommand{\Sec}[1]{Section~\ref{sec:#1}} 

\newcommand{\pt}[1]{\left(#1\right)} 

\newcommand{\jp}[1]{\begin{CJK}{UTF8}{ipxm}#1\end{CJK}} 

\newcommand{\drawfig}[4]{ 
    \begin{figure}[#1]
    \centering 
    \vspace{0mm}
    \includegraphics[width=#2]{#3} 
    \vspace{-4mm}
    \caption{#4}
    \label{fig:#3}
    \vspace{-4mm}
    \end{figure}
}

\begin{document}
%
\title{Building speech corpus with diverse voice characteristics for its prompt-based representation}
%
%
%

\author{Aya~Watanabe,~\IEEEmembership{Student~Member,~IEEE,}
        Shinnosuke~Takamichi,~\IEEEmembership{Member,~IEEE,}
        Yuki~Saito,~\IEEEmembership{Member,~IEEE}
        Wataru~Nakata,
        Detai~Xin,~\IEEEmembership{Student~Member,~IEEE,}
        and~Hiroshi~Saruwatari,~\IEEEmembership{Member,~IEEE}
\thanks{The authors are with the Graduate School of Information Science and Technology, The University of Tokyo, Bunkyo-ku, Tokyo 113-8656, Japan (e-mail: aya-watnabe@g.ecc.u-tokyo.ac.jp; shinnosuke\_takamichi@ipc.i.u-tokyo.ac.jp;  yuuki\_saito@ipc.i.u-tokyo.ac.jp; nakata-wataru855@g.ecc.u-tokyo.ac.jp; detai\_xin@ipc.i.u-tokyo.ac.jp; hiroshi\_saruwatari@ipc.i.u-tokyo.ac.jp).}
\thanks{Manuscript received April 19, 2005.}}

%
%

\markboth{Journal of \LaTeX\ Class Files,~Vol.~14, No.~8, August~2015}%
{Shell \MakeLowercase{\textit{et al.}}: Bare Demo of IEEEtran.cls for IEEE Journals}
%



\maketitle

\begin{abstract}
In text-to-speech synthesis, the ability to control voice characteristics is vital for various applications. By leveraging thriving text prompt-based generation techniques, it should be possible to enhance the nuanced control of voice characteristics. While previous research has explored the prompt-based manipulation of voice characteristics, most studies have used pre-recorded speech, which limits the diversity of voice characteristics available.
Thus, we aim to address this gap by creating a novel corpus and developing a model for prompt-based manipulation of voice characteristics in text-to-speech synthesis, facilitating a broader range of voice characteristics.
Specifically, we propose a method to build a sizable corpus pairing voice characteristics descriptions with corresponding speech samples. This involves automatically gathering voice-related speech data from the Internet, ensuring its quality, and manually annotating it using crowdsourcing. We implement this method with Japanese language data and analyze the results to validate its effectiveness.
Subsequently, we propose a construction method of the model to retrieve speech from voice characteristics descriptions based on a contrastive learning method. We train the model using not only conservative contrastive learning but also feature prediction learning to predict quantitative speech features corresponding to voice characteristics. We evaluate the model performance via experiments with the corpus we constructed above.

\end{abstract}

%
\IEEEpeerreviewmaketitle

\section{Introduction}
\label{sec:intro}

\IEEEPARstart{I}{n} human speech production, the speaker's voice conveys not only linguistic information but also non-linguistic cues. These include voice characteristics such as voice quality resulting from vocal tract shape and speaking styles reflecting emotions, speaking situations etc. Text-to-speech (TTS), the task to automatically imitate speech production, involves two significant challenges: synthesizing highly intelligible speech from the provided text and controlling the voice characteristics. In recent years, the introduction of deep neural network (DNN)-based methods~\cite{oord2016wavenet,wang2017tacotron,ren2019fastspeech} has led to tremendous improvements in terms of the first challenge, enabling the generation of highly faithful voices that closely resemble human speech. However, there is still room for improvement in controlling voice characteristics, which have a significant influence on the listener's perception and affect their understanding of the speaker's personality, emotion, and overall impression. Several elements for methods that control voice characteristics have been proposed, 
such as a speaker index~\cite{hojo16speakercode}, 
speaker attributes~\cite{stanton2022speaker,watanabe22mid-attribute-speaker-generation}, 
personality~\cite{gustafson21personality-in-the-mix}, 
and so on~\cite{zhang19learning-to-speak,rui21reinforcement-learning-emotional-tts,ohta10voice-quality-gmm-vc,raitio20controllable-tts}.
However, these methods are limited in that they only enable control over a narrow and simplistic range of voice characteristics.

These days, there has been significant advancement in techniques for media generation utilizing free-form text descriptions, commonly referred to as text prompts. This progress is evident in fields such as text-to-image~\cite{ramesh21dalle}, text-to-audio~\cite{elizalde22clap}, text-to-music~\cite{huang22mulan}, and text-to-video~\cite{ho22imagenvideo}. 
These kinds of prompt-based media generation are beneficial for controlling complicated media components, and they come with the advantage of large language models (LLMs)~\cite{radford21clip,elizalde22clap}, which are also improving these days thanks to using abundant amounts of web-crawled data~\cite{lin14ms-coco,sharma18conceptualcaptions,kim19audiocaps,drossos20clotho,wu22laion-audio}. Following these trends, free-form description-based voice characteristics control is poised to open new doors for TTS tasks. Hence, our goal is to develop TTS capable of controlling voice characteristics by free-form descriptions. Hereafter, we refer to this free-form description and TTS synthesizer as the ``voice characteristics descriptions'' and ``prompt-based TTS,'' respectively.

While there are already some methods of prompt-based TTS~\cite{guo22prompttts,yang23instructtts,zhang2023promptspeaker,lyth2024natural}, they use corpora based on the existing TTS corpora~\cite{zen19libritts,takamichi2020jsut}, which tend to cover only a limited range of voice characteristics. Given the significant advancements in prompt-based generation facilitated by LLMs in other domains, the corpus for prompt-based TTS should ideally encompass a wide range of voice characteristics, but neither an open corpus nor a scalable methodology to construct such a corpus is currently available (see \Table{corpus_compare}).

\drawfig{t}{\linewidth}{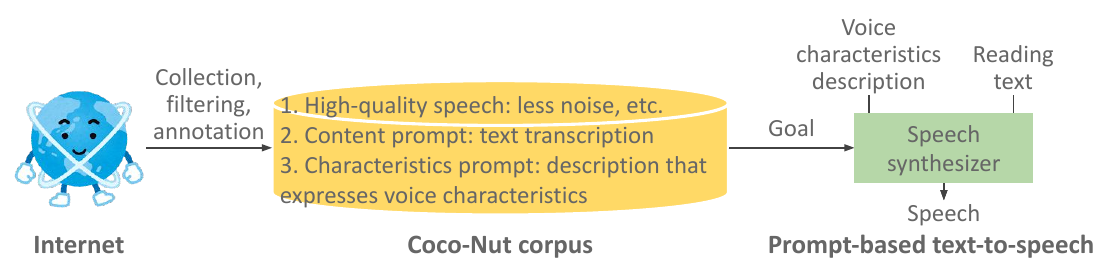}
{Our Coco-Nut corpus towards prompt-based TTS. Voice characteristics description and reading text are, for example, ``middle-aged man's voice speaking in a clear and polite tone'' and ``Welcome to our office!'', respectively. A speech synthesizer synthesizes the speech of the prompted content on the basis of the prompted voice characteristics.}

\begin{table}[t]
\centering 
\caption{Comparison of corpora of speech-voice characteristics description.}
\begin{tabular}{r|c c}
        & Data source                               & Public corpus \\ \hline
    Guo et al.~\cite{guo22prompttts}       
        & Public TTS corpus~\cite{zen19libritts}    & \checkmark\\
    Yang et al.~\cite{yang23instructtts}    
        & Internal TTS corpus                       & –         \\
    Zhang et al.~\cite{zhang2023promptspeaker} 
        & Internal TTS corpus                       & –         \\ \hline
    \textbf{Ours} 
        & Internet (diverse data)                   & \checkmark
\end{tabular}
\label{tab:corpus_compare}
\end{table}

In this paper, we propose a methodology for constructing a corpus toward prompt-based TTS featuring plentiful voice characteristics by utilizing in-the-wild speech data from the Internet. Our methodology consists of three main steps: 1) automatic collection of voice-related audio data from the Internet, 2) quality assurance to enhance both the linguistic and acoustic quality of speech in the corpus, and 3) manual annotation of voice characteristics descriptions using crowdsourcing. With this methodology, we construct an open corpus, \textit{Coco-Nut}\footnote{\textbf{Co}rpus of \textbf{co}nnecting \textbf{N}ihongo \textbf{u}tterance and \textbf{t}ext. ``Nihongo'' means ``the Japanese language'' in Japanese.}, which is available on our project page\footnote{\url{https://sites.google.com/site/shinnosuketakamichi/research-topics/coconut_corpus}}. 
\Fig{figure/overview.pdf} shows the total process from corpus construction to synthesizing speech that aligns with the prompted linguistic content and voice characteristics with our corpus.

We also propose a method to construct a retrieval model to encode speech and voice characteristics descriptions into embedding and retrieve speech from descriptions by using our in-the-wild corpus. The model is trained based on contrastive learning~\cite{radford21clip,elizalde22clap} and a new objective to predict features related to voice characteristics. This augmentation is aimed at improving the performance of embedding. Experimental evaluations on the Coco-Nut corpus reveal the effectiveness of this training method.

The contributions of this work can be summarized as follows.
\begin{itemize}
    \item We propose a method to construct a paired corpus of speech and voice characteristics descriptions with a broad range of voice characteristics from the Internet. The proposed method includes data collection, quality assurance, and annotation.
    \item  We open-sourced the corpus, which is the first corpus that covers diverse in-the-wild voice characteristics.
    \item We propose a training method for aligning speech and voice characteristics descriptions.
\end{itemize}
 The construction method and some analysis of the corpus have previously been published in an international conference paper~\cite{watanabe2023coconut}. In the current paper, we provide a deeper analysis of the corpus and present a novel training for aligning the speech and voice characteristics descriptions. 
\section{Related work}
\label{sec:related-work}

\vspace{-0mm}
\subsection{Sequence generation from text}  \vspace{-0mm} \label{sec:related-work_text-representation}
In contrast to generating images, which lack a temporal dimension, generating sequential data such as text-to-video and text-to-audio requires a consideration of the following aspects: 1) \textit{overall concepts} that represent characteristics of the entire sequence and 2) \textit{sequence concepts} that represent characteristics of changes in the sequence. 

There are two approaches for describing these concepts by free-form texts. 
The first is to put both concepts into a single text, e.g., ``wooden figurine surfing on a surfboard in space''~\cite{ho22imagenvideo} in text-to-video and ``hip-hop features rap with an electronic backing''~\cite{huang22mulan} in text-to-music. This method is well-suited for applications that generate sequences based on general descriptions~\footnote{For enhancement, MusicLM~\cite{agostinelli23musiclm} switches the description at fixed intervals (as noted in the paper, every 15 seconds). This adjustment enables finer control over the changes in the generated sequences.}. The second is to describe each concept separately. Examples of this include ``bat hitting'' (overall concept) and ``ki-i-i-n'' (sequence concept) in text-to-audio~\cite{ohnaka22visual-onoma-to-wave}\footnote{In LAION-Audio-630K~\cite{laion-ai-audio}, the approach involves overall concepts for non-speech environmental sounds and sequence concepts (transcriptions) for speech-containing sounds.} and ``A toy fireman is lifting weights'' (sequence concept) in text-to-video~\cite{molad23dreamix}\footnote{In the paper, overall concept is given by an image.}. These methods are well-suited for applications demanding precise control over sequences, such as TTS, where linguistic content and voice characteristics are frequently managed independently~\cite{guo22prompttts,yang23instructtts,zhang2023promptspeaker}. Therefore, prompt-based TTS utilizes corpora annotated with voice characteristics descriptions independent of transcriptions.

\subsection{Dataset for text-to-image}  \vspace{-0mm}\label{sec:related-work_text-image}
Text-to-image models require pairs of an image and text prompt that describes the image content for training them.
DALL-E~\cite{ramesh21dalle}, known as a pioneer in text-to-image, is trained using the image captioning set of the MS-COCO dataset~\cite{lin14ms-coco} and web-crawled data~\cite{sharma18conceptualcaptions}. 
MS-COCO is a dataset of manually annotated images including texts that describe the image content (called captions), used for image captioning. In addition to MS-COCO, using a wide range of data from the Internet in training improves the image diversity of synthesis~\cite{ramesh21dalle}. Although HTML images and their accompanying alt-tag texts are remarkably beneficial in terms of providing plenty of text-image pairs, data filtering is required due to the noisiness of the Internet data. For data filtering, contrastive learning (see \Sec{contrastive-learning})-based models such as CLIP~\cite{radford21clip} are often used. Following the success in the image field, data diversity and contrastive learning are presumably important in other domains as well, e.g., voice characteristics in this paper.

\vspace{-0mm}
\subsection{Dataset for text-to-audio and text-to-music}  \vspace{-0mm}\label{sec:related-work_text-audio}
As with images, datasets for captioning are also available for text-to-audio.
Typical examples are AudioCaps~\cite{kim19audiocaps} and Clotho~\cite{drossos20clotho}. Additionally, the text-audio version of CLIP, CLAP~\cite{elizalde22clap}, is also used for data filtering~\cite{wu22laion-audio} before the training. In text-to-music, MuLan~\cite{huang22mulan} introduces a method for extracting music videos from the Internet and developing a machine learning model to determine if the accompanying text accurately describes the music. This approach can potentially be extended to other forms of acoustic media beyond music.

Unlike the text-to-audio and text-to-music cases, datasets for prompt-based TTS are very limited\footnote{Audio captioning datasets~\cite{kim19audiocaps,drossos20clotho} include human voices as an environmental sound, but the voices do not strongly specify linguistic content.}. 
In many existing methodologies, corpora utilized for training consist of resources tailored for TTS applications, such as LibriTTS~\cite{zen19libritts}, with manual annotation of voice characteristics descriptions~\cite{guo22prompttts}, or internally collected speech recorded from a small number of speakers~\cite{yang23instructtts}. However, these approaches suffer from several limitations, including a lack of diversity in voice characteristics and the prevalence of non-public corpora.
Considering the contribution of Internet data discussed in \Sec{related-work_text-image}, it is necessary to establish a methodology of corpus construction from Internet data. Also, the accessibility of the corpus is important.

\subsection{Contrastive learning for text-audio}
\label{sec:contrastive-learning}
\drawfig{t}{0.9\linewidth}{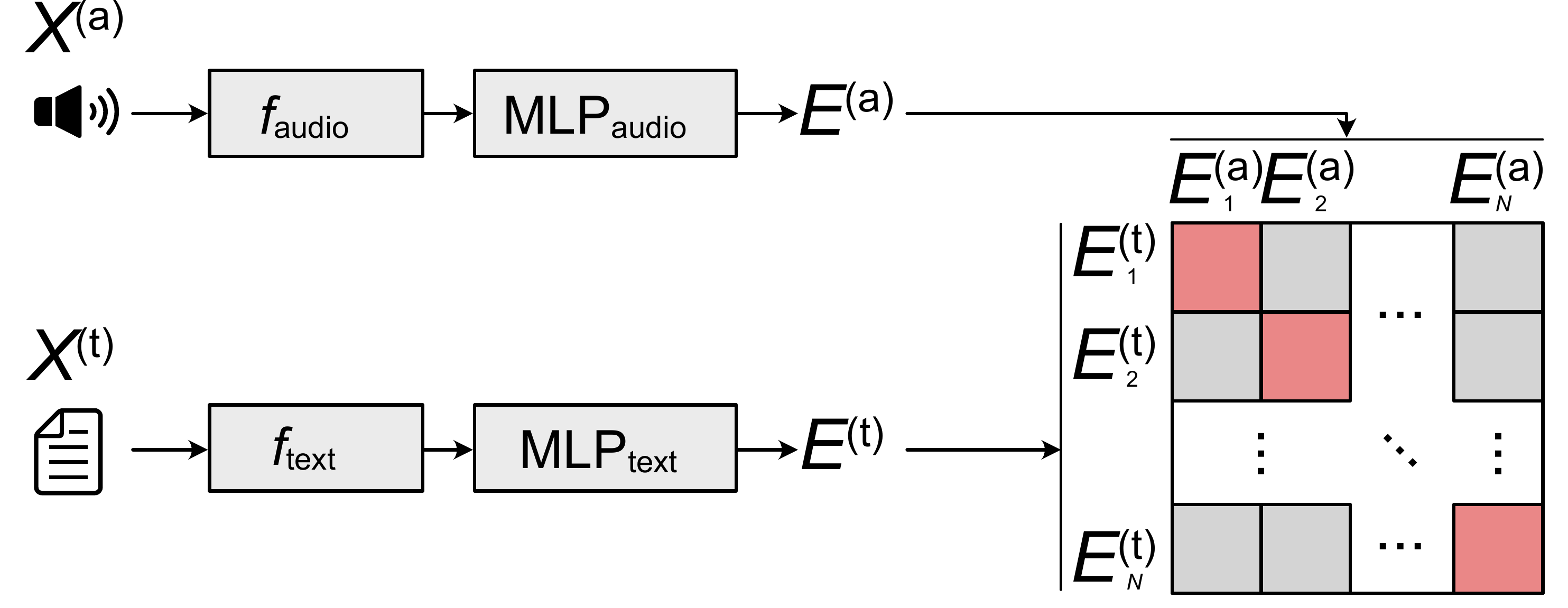}{CLAP overview. Embedding vectors from text and audio are learned by contrastive learning.}
Contrastive learning, exemplified by CLIP~\cite{radford21clip}, aims to learn correspondences between different forms of media and embed them into the same space. An example of this is CLAP~\cite{elizalde22clap}, which conducts contrastive learning for environmental sounds and their captions. An overview of the structure of CLAP is depicted in the top row of \Fig{figure/clap.pdf}.
CLAP encodes the waveform $X^{(\text{a})}$ and its corresponding caption $X^{(\text{t})}$ using pretrained encoders $f_{\text{audio}}(\cdot)$ and $f_{\text{text}}(\cdot)$, respectively, transforming them into vectors $E^{(\text{a})}$ and $E^{(\text{t})}$ of the same dimensionality using multi-layer perceptrons $\text{MLP}_{\text{audio}}(\cdot)$ and $\text{MLP}_{\text{text}}(\cdot)$. During training, utilizing pairs of waveforms and captions $(X^{(\text{a})}_i, X^{(\text{t})}_i)$, where $i$ denotes the $i$th pair, the model is trained by minimizing the following loss function:
\begin{align}
    L_{\rm{CLAP}} &=  \frac{1}{2N} \sum_{i=1}^N \left(
        L_{{(\rm{a} \to \rm{t})} i} + L_{{(\rm{t} \to \rm{a})} i} 
    \right), \label{eq:clap}
\end{align}
where
\begin{align}
    L_{({\rm a \to t})i} &= 
        \log \frac{\exp\pt{E^{(\rm{a})}_i \cdot E^{(\rm{t})}_i / \tau}}
            {\sum_{j=1}^N \exp\pt{E^{(\rm{a})}_i \cdot E^{(\rm{t})}_j / \tau}}, \\
    L_{({\rm t \to a})i} &=
        \log \frac{\exp\pt{E^{(\rm{t})}_i \cdot E^{(\rm{a})}_i / \tau}}
            {\sum_{j=1}^N \exp\pt{E^{(\rm{t})}_i \cdot E^{(\rm{a})}_j / \tau}}.
\end{align}

Here, $N$ denotes the total number of training pairs used in a single loss computation (or batch size in mini-batch training), and $\tau$ represents the temperature parameter.
Through training, positive pairs tend to be embedded closer to each other.

CLAP was originally proposed for environmental sounds, but it can be extended to other tasks, e.g., speech and voice characteristics descriptions~\cite{nagrani2020voxceleb,maekawa2003csj}. Also, additional training objectives can enhance the performance of the contrastive learning models~\cite{yeh2023flap}  
\section{Corpus construction}
\label{sec:corpus-construction}

\drawfig{t}{0.99\linewidth}{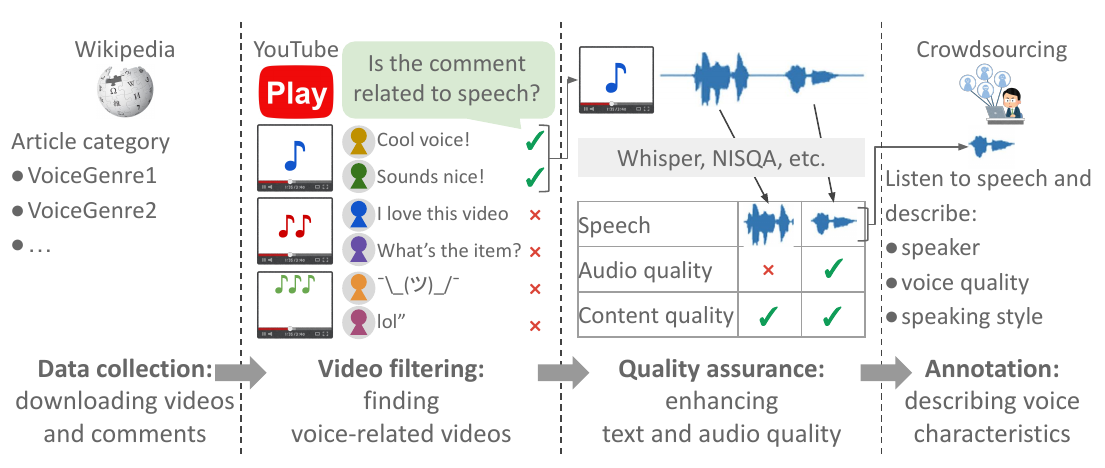}
{Procedure of corpus construction.}

The corpus for prompt-based TTS should include:
\begin{enumerate} \vspace{-0mm}
\item \textbf{High-quality speech.} Speech data suitable for TTS. Unlike data in speech-to-text (automatic speech recognition) corpora~\cite{chen2021gigaspeech,takamichi21jtubespeech,yin23reazonspeech}, it should be high-quality, e.g., contain less noise.
\item \textbf{Transcription.} Text transcriptions of speech, corresponding to the ``sequence concept'' described in \Sec{related-work_text-representation}.
\item \textbf{Voice characteristics description.} Free-form descriptions that express the characteristics of speech, corresponding to the ``overall concept'' described in \Sec{related-work_text-representation}.
\end{enumerate} \vspace{-0mm}
As mentioned in \Sec{related-work_text-audio}, conventional prompt-based TTS methods~\cite{guo22prompttts,yang23instructtts,zhang2023promptspeaker,lyth2024natural} use corpora based on existing TTS corpora including high-quality speech and their transcriptions \cite{takamichi2020jsut,zen19libritts}, with additional annotation of voice characteristics descriptions. Since such corpora often lack diversity of voice characteristics, we propose a method that builds a corpus from very noisy but diverse Internet data.

It consists of the following four steps following below, as illustrated in \Fig{figure/procedure.pdf}.
\begin{enumerate}
\item \textbf{Data collection.} Speech data candidates, in the form of videos, are searched out and downloaded from the Internet. 
\item \textbf{Video filtering.} Impressive voice data are filtered from the candidates. Here, ``impressive voice'' refers to voices those that have received a sufficiently large number of responses on the Internet. Such data are expected to be characteristic voices, which will help improve the diversity of the corpus and can easily be annotated with voice characteristics descriptions.
\item \textbf{Quality assurance.} Speech are further filtered to guarantee the quality of the corpus, in both linguistic and acoustic aspects. 
\item \textbf{Manual annotation.} Voice characteristics descriptions are manually annotated to the speech data.
\end{enumerate} 
While the demonstration in \Sec{corpus-analysis} focuses on Japanese, it is important to note that the method itself is applicable to languages other than Japanese.

The subsequent subsections explain these steps in detail.

\subsection{Data collection} \label{sec:proposed-method-data-collection}
To collect speech data candidates, we make search keywords and phrases and input them to the search engines of video-sharing platforms such as YouTube\footnote{\url{https://www.youtube.com}}. From Wikipedia\footnote{\url{https://en.wikipedia.org} in English.}, we choose article categories related to speech\footnote{For example, \url{https://en.wikipedia.org/wiki/List_of_YouTubers} in English.} in the target language, utilizing the titles of Wikipedia articles within those categories as search keywords. Additionally, we make search phrases by concatenating the collected keywords with relevant phrases (e.g., ``[article title] short clip''). Upon conducting the search, we obtain audio data extracted from found videos and, in addition, various metadata of videos e.g., video IDs, video titles, and viewer comments.   

\subsection{Video filtering}
After collecting the video data as described above, we filter the data to acquire videos featuring ``impressive voices.'' We focus on videos that have a significant number of viewer comments on the speech heard in the videos. This filtering process involves two stages described below to ensure precision.

\begin{enumerate} 
\item \textbf{Keyword matching-based pre-filtering.} 
Some collected videos may lack distinctive voices or may not contain any speech at all. To address this, we implement a rule-based video filter. This filter relies on a predefined set of keywords associated with voice characteristics (e.g., ``listen'') to assess whether viewer comments mention these keywords. If the number of comments containing these keywords for a particular video surpasses a specified threshold, the video is adopted.

\item \textbf{Machine learning-based filtering.}
After filtering by keyword-matching, machine learning-based speech-related comment classification is utilized for further detailed filtering. We create a small dataset for training the classifier. We randomly extract some viewer comments from videos and annotate them by using crowdsourcing. A pair of comment and title of the video are shown to the crowdworkers\footnote{For example, ``Video title: My daily voice training method. Comment: Cool Voice!'' The correct answer is ``1) related to speaking voice.'' Presenting the title helps crowdworkers annotate by giving the imagination of the content of the video.}, who then answer whether the shown comment is 1) related to speaking voice, 2) related to singing voice, or 3) others. 
Before annotating, we instruct crowdworkers that ``1)'' encompasses comments discussing voice characteristics but excludes comments about what is said.

Then, the speech-related comment classifier is trained with annotated data. This model is based on BERT~\cite{devlin2018bert} followed by a linear layer. The input comprises a video title and comment concatenated with a ``[SEP]'' token, indicating sentence separation in BERT. The output is binary, categorizing comments into either whether speech-related or any other labels (including 2) related to singing voice and 3) others).
To enhance the classification performance, we utilize the auxiliary keywords. We re-utilize the keyword set that was used in the keyword matching-based filtering to make a subset of keywords and restrict training and evaluation of the classifier to comments that match one of the words in the subset.

\end{enumerate} \vspace{-0mm}

Subsequently, the voices from the filtered videos proceed to the next stage of ``quality assurance.''

\subsection{Quality assurance} \vspace{-0mm}
Since we collect data from the Internet, there are some speech samples of low quality that would be difficult to use. To ensure the quality of speech in the constructed corpus from the viewpoints of of both acoustic and linguistic quality, we conduct the following processes.

\vspace{-0mm}
\subsubsection{Audio quality} \vspace{-0mm}
To ensure the acoustic quality of speech, the following operations are performed.
\begin{enumerate}
\item \textbf{Voice activity detection (VAD).}
VAD is performed to extract only the segments of speech from the entire audio of video. We use inaSpeechSegmenter~\cite{ddoukhanicassp2018} to detect speech segments from the whole video. As a result, each video is processed into multiple segmented speech segments.

\item
\textbf{Denoising.} We use Demucs\footnote{\url{https://github.com/facebookresearch/demucs}} source separation model to extract the voices from noise-contaminated segments.

\item
\textbf{Audio quality assessment.} The speech obtained through web crawling have variations in audio quality due to differences in recording device quality and effective frequency band. Additionally, the denoising process may occasionally suppress speech components. To assess the degradation in quality resulting from these factors, we utilize NISQA~\cite{mittag21nisqa}, a multidimensional speech quality predictor. NISQA scores are computed for each speech segment, and segments with scores below a predetermined threshold are filtered out\footnote{We found that speech component drop can be evaluated by the NISQA score.}.

\item
\textbf{Threshold for duration and audio volume.}
We establish acceptable duration ranges to exclude excessively long or short speech segments. Additionally, we set a volume threshold to filter out inaudible (low-volume) speech.

\item 
\textbf{Detection of multi-speaker voice and singing voice.}
We manually exclude data not intended for TTS, specifically singing voices and multi-speaker voices (including short nods).

\item 
\textbf{Voice characteristics variation.} 
To ensure the corpus encompasses a diverse range of voice characteristics, we implement hierarchical clustering utilizing Ward's method~\cite{ward63clustering} based on the distances of $x$-vectors~\cite{snyder2018x}. These vectors not only reflect voice quality but also speech style, as suggested by~\cite{brown2021playing}. $x$-vectors are extracted for each speech segment using a pretrained $x$-vector extractor. Given that speech segments with similar voice characteristics are anticipated to be grouped together, we randomly select one speech segment to serve as the representative of each cluster.
\end{enumerate} \vspace{-0mm}

\subsubsection{Content quality}  \vspace{-0mm}
To ensure the linguistic quality of speech, the following operations are performed.

\begin{enumerate} 
\item 
\textbf{Speech-to-text and language identification.}
To transcribe speech, we utilize pre-trained Whisper speech-to-text models~\cite{radford2022robust}. Concurrently, we identify the language of the speech by Whisper and filter out any speech not in the target language. Additionally, manual language identification is conducted to further enhance the corpus quality\footnote{We observed that relying solely on Whisper for language identification would result in the inclusion of many non-target language voices.}.

\item
\textbf{NSFW (not safe for work) word detection.} 
We exclude transcriptions containing NSFW words. This is achieved through keyword matching-based NSFW word detection, where the text is filtered out if the lemmatized word is found in the NSFW word dictionary. Additionally, further manual detection is carried out to improve the quality of the corpus.

\item
\textbf{Non-verbal voice detection.}
Since TTS systems typically do not handle non-verbal voices, such as screams, we filter them out using an LLM on transcriptions. Masked language model (MLM) scores~\cite{salazar20maskedlaugagemodel} based on BERT~\cite{devlin2018bert} are computed for each segment's transcription. Given that the masked tokens of transcriptions are highly predictable from the adjacent tokens\footnote{For example, with ``aa[MASK]aaaa,'' a partially masked content prompt of a scream, it can easily be predicted that ``[MASK]'' will be ``aa.''}, the MLM score of the non-verbal voice tends to be higher. We manually set a threshold against the MLM score and filter out speech with a score exceeding this threshold.
\end{enumerate} 

\vspace{-0mm}
\subsection{Manual annotation} \vspace{-0mm}
Finally, we utilize crowdsourcing to add voice characteristics descriptions to the collected speech. Crowdworkers listen to the presented speech and describe its voice characteristics, including speaker attributes, voice quality, and speaking style (e.g., angry, fast) in a free-form description within some instructional parameters\footnote{The actual English-translated instruction is ``Describe the kind of speaker (age, gender, etc.), voice quality (brisk, low, etc.), and speaking style (angry, fast, etc.) in a free-form description of at least $20$ characters. Do not include the linguistic content of the speech, and do not use expressions that indicate personal likes and dislikes (e.g., my favorite voice or disliked way of speaking).''. }.

After collecting the voice characteristics descriptions, we manually filtered out ones including proper nouns and persons' name, e.g., ``The voice is similar to [celebrity's real name].'' We undertake these steps to prevent models trained on this corpus from generating specific individuals' speech by inputting their name. Additionally, we perform text normalization to clean up the descriptions.

\section{Corpus construction settings and results}
\label{sec:corpus-analysis}
Using the process outlined in \Sec{corpus-construction}, we constructed the Coco-Nut corpus. In this section, we provide detailed descriptions of the construction conditions and analyze the developed corpus.

\subsection{Data collection}
\label{sec:experiments-collection}
The target language was Japanese. The data collection was conducted from July 2022 to March 2023. The number of downloaded comments per video was limited to the top $100$ comments, sorted in descending order by the number of ``Likes.'' After extracting comments only in the target language using rule-based language identification, only comments with more than $3$ and less than $50$ characters were retained. \Table{data-collection} shows the results of the data collection.

\begin{table}[t]
\centering 
\caption{Results of data collection} \footnotesize
\begin{tabular}{l|l}
Retrieved item               & Value          \\ \hline
No. of article-categories                 & $180$ \\
No. of search-phrases             & $0.10$M \\
No. of videos found in the search & $1.14$M  \\
Audio duration               & $0.30$M hours \\
No. of comments                   & $24.2$M \\
\end{tabular}
\label{tab:data-collection}
\vspace{-4mm}
\end{table}

\subsection{Filtering} \label{sec:experiments-filtering}
We used eight words for keyword matching-based pre-filtering: ``\jp{声}'', ``\jp{ボイス} '', ``\jp{ヴォイス}'' (three versions of voice), ``\jp{響}'' (resonance), ``\jp{音}'' (sound), ``\jp{聴}'' (listen), ``\jp{聞}'' (hear), and ``\jp{歌}'' (song).
We set the threshold for the number of keyword-matching comments per video to $10$.

We utilized pre-trained BERT~\cite{devlin2018bert} model\footnote{\url{https://huggingface.co/cl-tohoku/bert-base-japanese}} for machine learning-based filtering. 
For fine-tuning, we collected $32{,}453$ labels for comments, out of which $11{,}647$ were ``speech-related.'' A total of $80$\% of labels were used for training and $20$\% for evaluation.
To determine the best selection of subset, we evaluated the model performance on all combinations of the subsets.
Finally, seven different subsets were selected based on high precision of classifiers, as we needed to exclude non-desirable comments rather than to prevent accurate comments from being excluded. The average precision of the seven classifiers was $54.3$\%. Given that the classifier without the keyword subsets performed with the average precision of just $38.6$\%, we conclude it is effective to use the keyword subsets.
We classified unlabeled comments with trained classifiers. Videos were retained if they had $10$ or more comments classified as ``speech-related'' by any of the seven classifiers.
Hereafter, $1{,}523$ videos were used for further processing. These are the videos remaining from the subset collected in \Sec{experiments-collection}, after the process outlined in this subsection.

\subsection{Quality assurance}
After implementing VAD, we obtained $54{,}610$ speech segments from $1{,}523$ videos.
\Fig{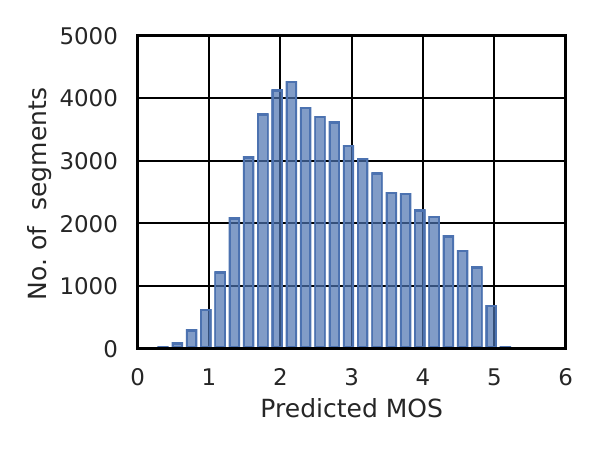} shows the distribution of predicted mean opinion scores (MOSs) on audio quality by NISQA. We set the threshold to $2$ because it is the most frequent score.
Also, only segments with a duration between $2$ and $10$ seconds were retained.
We used Pydub\footnote{\url{https://github.com/jiaaro/pydub}} to check the audio volume, and segments with a volume of $-55$ dB or lower were excluded.

For transcription, we utilized both the tiny and large models of Whisper~\cite{radford2022robust}, as the former tends to prioritize fidelity to the speech while the latter prioritizes grammatical correctness\footnote{The average word error rate (WER) of transcriptions from the Whisper large model was 22.1\%. Upon final publication, we corrected transcriptions manually to ensure a WER of 0\%.}.

The NSFW detection was performed with MeCab\footnote{\url{https://taku910.github.io/mecab/}} and the Japanese NSFW dictionary\footnote{\url{https://github.com/MosasoM/inappropriate-words-ja}}.

We set the MLM score threshold for non-verbal voice detection to $-0.01$. \Fig{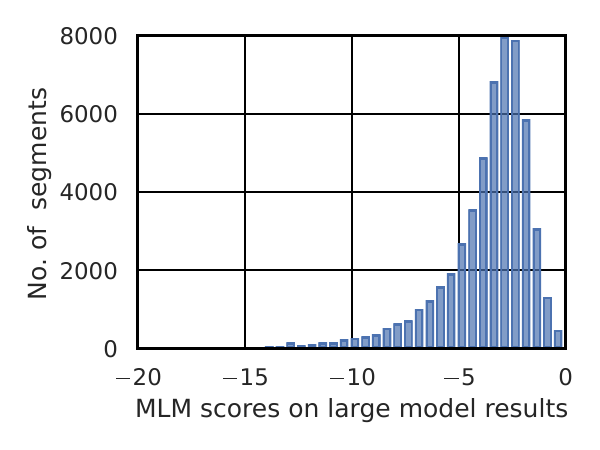} shows the distribution of MLM scores calculated by the transcriptions from the Whisper large model. According to this analysis, the MLM scores of collected segments are distributed around the peak of $-3$, with a range of roughly $\pm 2$ intervals. 
We observe that the percentage of segments whose MLM scores exceed $-0.01$ was approximately $0.05$\%, which is extremely low frequency.

\begin{figure}[t]
\centering
\begin{minipage}[t]{0.48\linewidth}
\centering
\includegraphics[width=0.98\linewidth]{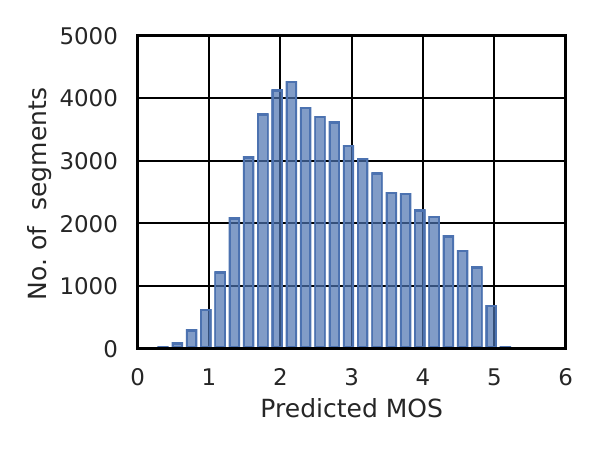}
\caption{Histogram of NISQA-predicted MOS on speech quality.}
\label{fig:figure/MOS_pred_hist.pdf}
\end{minipage}
\hfill
\begin{minipage}[t]{0.48\linewidth}
\centering
\includegraphics[width=0.98\linewidth]{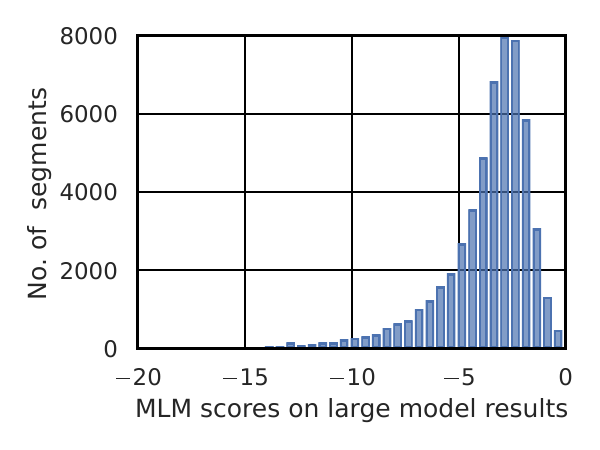}
\caption{Histogram of MLM scores.}
\label{fig:figure/MLM_hist.pdf}
\end{minipage}
\end{figure}

We utilized the $x$-vectors obtained from xvector\_jtubespeech\footnote{\url{https://github.com/sarulab-speech/xvector_jtubespeech}} for analyzing variations in voice characteristics. Hierarchical clustering was performed, resulting in the creation of $11{,}000$ clusters based on similarities in voice characteristics. A single speech segment was randomly chosen from each cluster. Subsequently, manual annotation was carried out through crowdsourcing to identify segments containing NSFW words, non-target language, and multiple speakers. Ultimately, a total of $7{,}667$ segments, amounting to $30{,}661$ seconds in length, were selected for further analysis.

\subsection{Annotation} 
We hired workers through the Lancers\footnote{\url{https://www.lancers.jp}}. Each worker annotated $10$ segments. There were a total of $1{,}318$ workers, and each was paid $200$ yen as reward.

Before annotation, for the machine learning usage, we designed the training, validation, and test sets to prevent data leakage stemming from similar voice characteristics within the same video or YouTube channel, and we ensured that the sets had no overlap in YouTube channels and comprised a varied selection of segments. Consequently, we acquired training, validation, and test sets containing $6{,}463$, $593$, and $611$ segments, respectively\footnote{We conducted quality evaluations by professional annotators, leading to the exclusion of several segments for public distributed version.}.
We designed our corpus to incorporate variations introduced by workers. In particular, for the training set, we included one voice characteristics description per segment, while for the other sets, we included five descriptions per segment, following previous studies~\cite{drossos20clotho, audiocaps}.

\subsection{Corpus analysis}

\begin{figure}[t]
\centering
\begin{minipage}[b]{0.52\linewidth}
\centering
\includegraphics[width=0.98\linewidth]{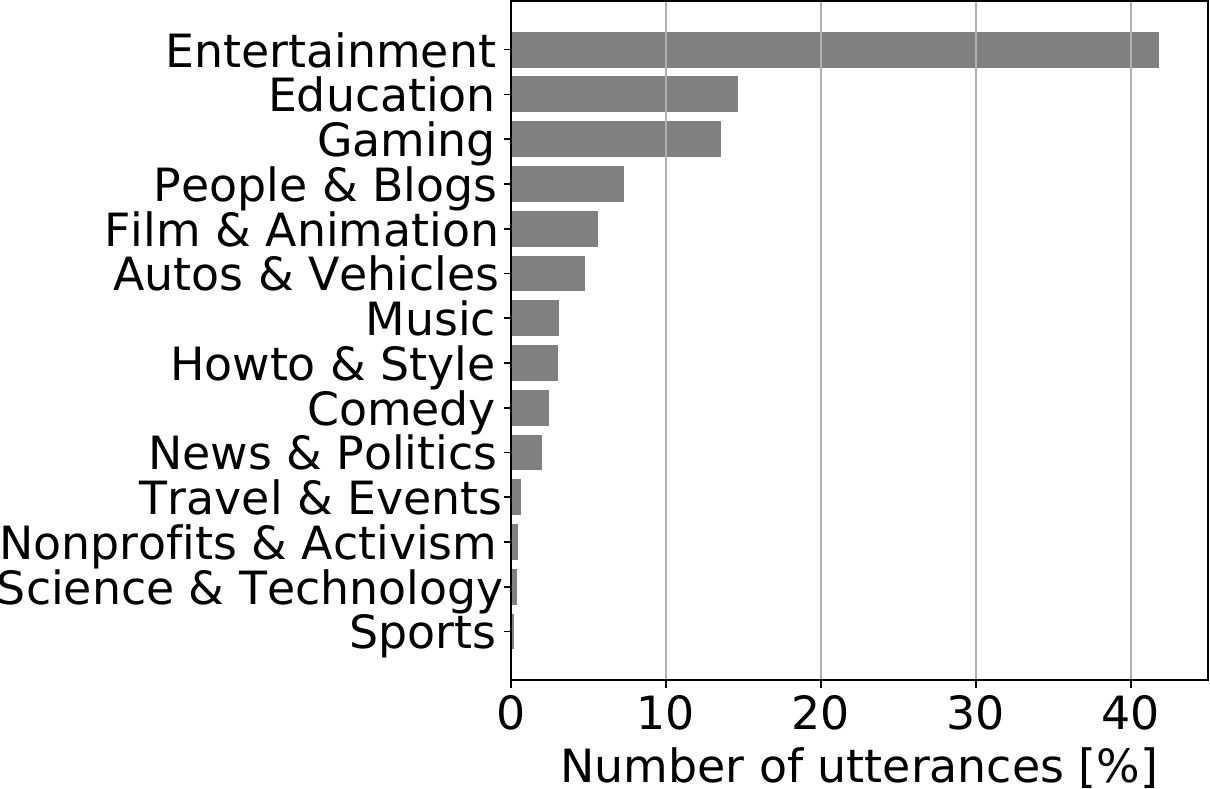}
\caption{Categories of speech.}
\label{fig:figure/video_category.pdf}
\end{minipage}
\begin{minipage}[b]{0.44\linewidth}
\centering
\includegraphics[width=0.98\linewidth]{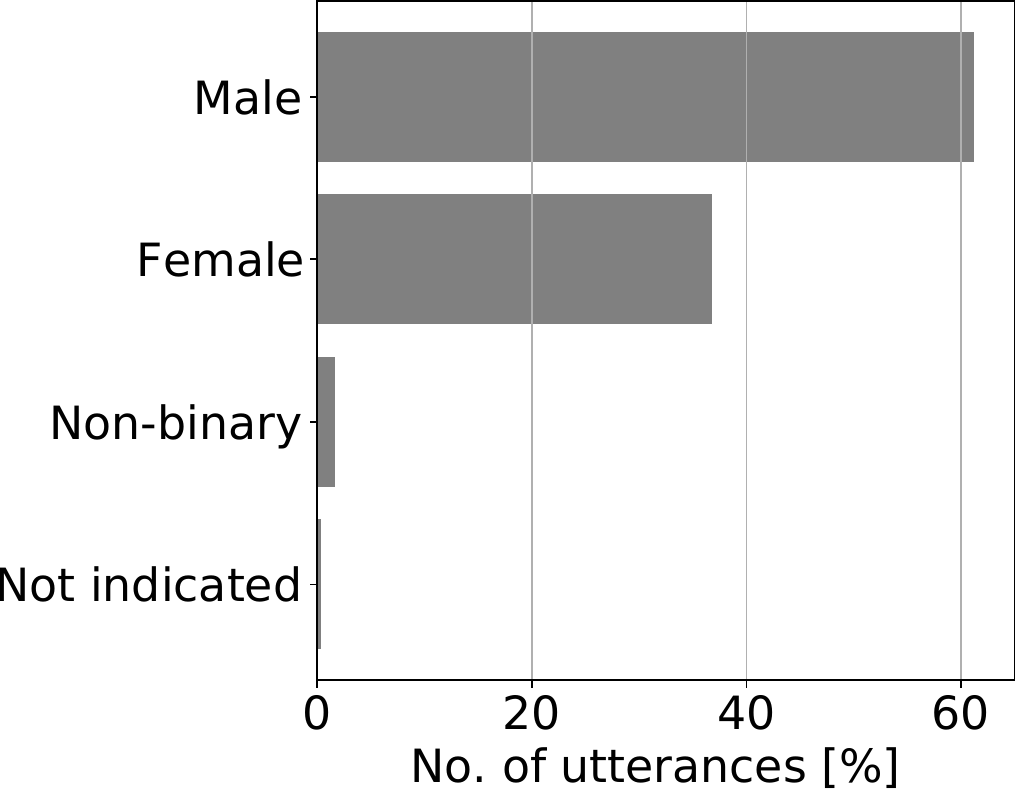}
\caption{Gender of speech.}
\label{fig:figure/gender.pdf}
\end{minipage}
\end{figure}

\begin{figure}[t]
\centering
\begin{minipage}[b]{0.43\linewidth}
\centering
\includegraphics[width=0.68\linewidth]{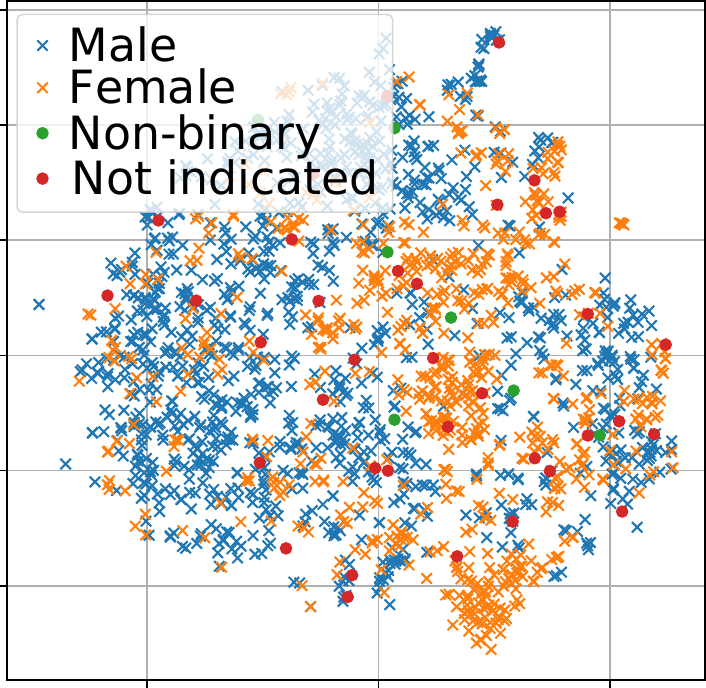}
\caption{$x$-vector distributions colored by gender.}
\label{fig:figure/xvector-gender.pdf}
\end{minipage}
\begin{minipage}[b]{0.55\linewidth}
\centering
\includegraphics[width=0.98\linewidth]{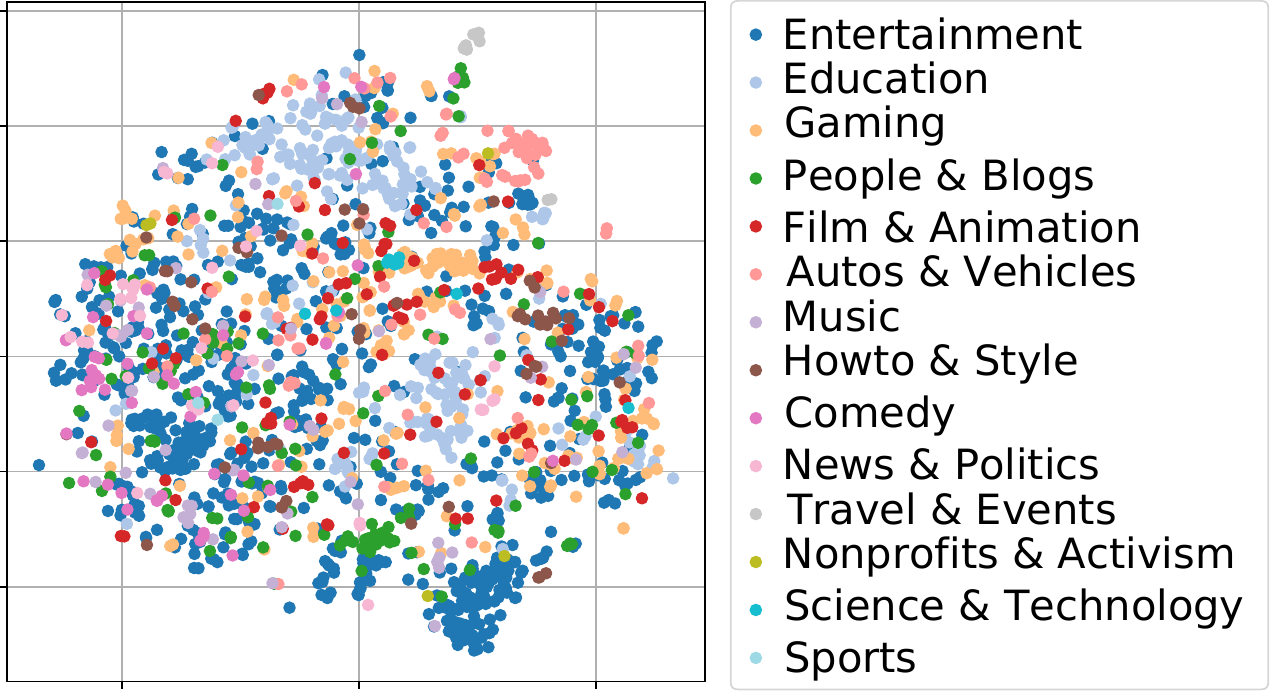}
\caption{$x$-vector distributions colored by category.}
\label{fig:figure/xvector-category.pdf}
\end{minipage}
\end{figure}

We analyze the constructed corpus focusing primarily on the diversity of data, which is the main purpose of the corpus, and examine the features of the collected voice characteristics descriptions.

\subsubsection{Video categories}
We examined the video category to which speech segments in the corpus belonged. Each segment's source video was categorized based on YouTube video categories. The results are shown in \Fig{figure/video_category.pdf}. The corpus encompasses $14$ categories, indicating a broad coverage. The top three categories (Entertainment, Education and Gaming) constitute roughly $70$\%, while minor categories such as Science \& Technology are also represented.

\subsubsection{Gender distributions}
We conducted automatic annotation of gender based on the voice characteristics descriptions to examine gender diversity. We labeled the descriptions based on the presence of gender-indicating characters (``\jp{男}'' for male, ``\jp{女}'' for female). In this labeling scheme, if only one gender-indicating character is present, the label will be assigned as male or female accordingly. If both are present, the label will be non-binary, and if neither is present, the label will be not-indicated.
\Fig{figure/gender.pdf} depicts the distribution of gender. While the majority of characteristic prompts are labeled as male or female, instances of non-binary and prompts not indicating gender (not-indicated) also exist. Similar to a typical TTS corpus, distinct clusters can be observed for male and female voices. However, non-binary and not-indicated categories do not form distinct clusters but are scattered throughout.
For a more detailed analysis, we provide a t-SNE visualization of $x$-vectors colored by gender in \Fig{figure/xvector-gender.pdf}. Clusters representing male and female voices are discernible, while the non-binary and not-indicated categories do not form clusters and are instead dispersed.
\label{sec:linguistic_analysis}
\begin{figure*}[t]
    \centering
    \begin{minipage}[t]{0.32\linewidth}
        \centering
        \includegraphics[width=0.98\linewidth]{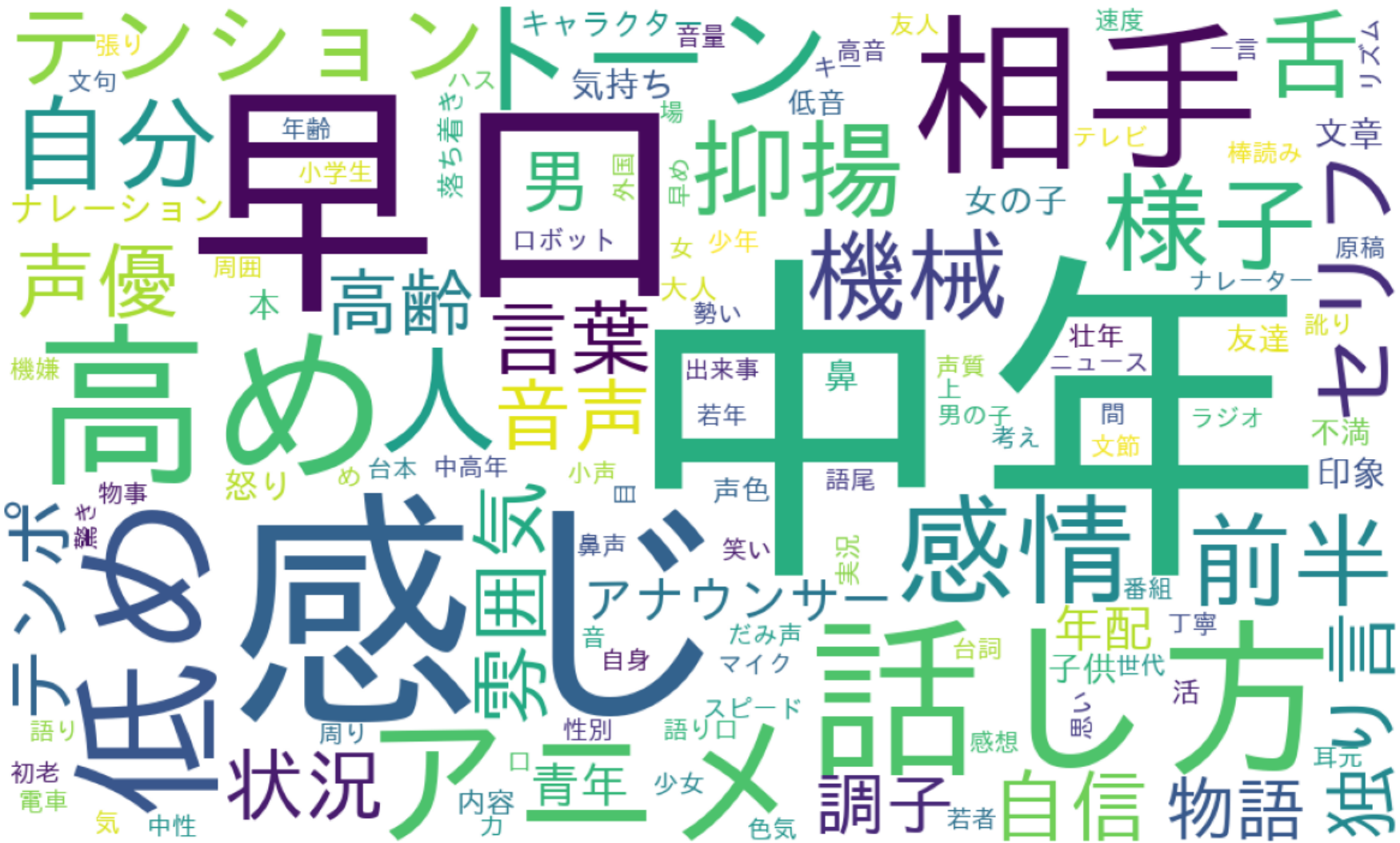}
    \end{minipage}
    \hfill
    \begin{minipage}[t]{0.32\linewidth}
        \centering
        \includegraphics[width=0.98\linewidth]{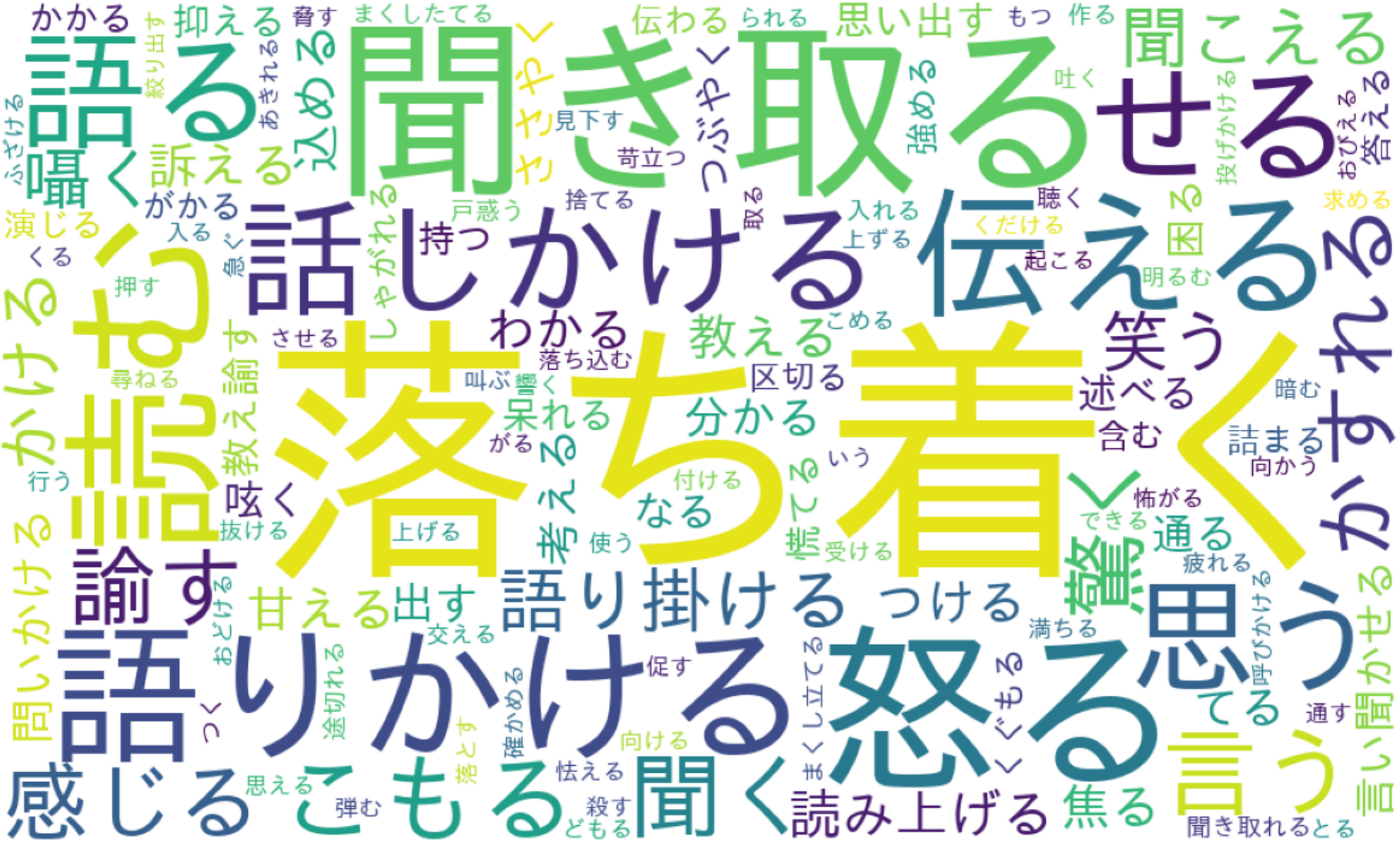}
    \end{minipage}
    \hfill
    \begin{minipage}[t]{0.32\linewidth}
        \centering
        \includegraphics[width=0.98\linewidth]{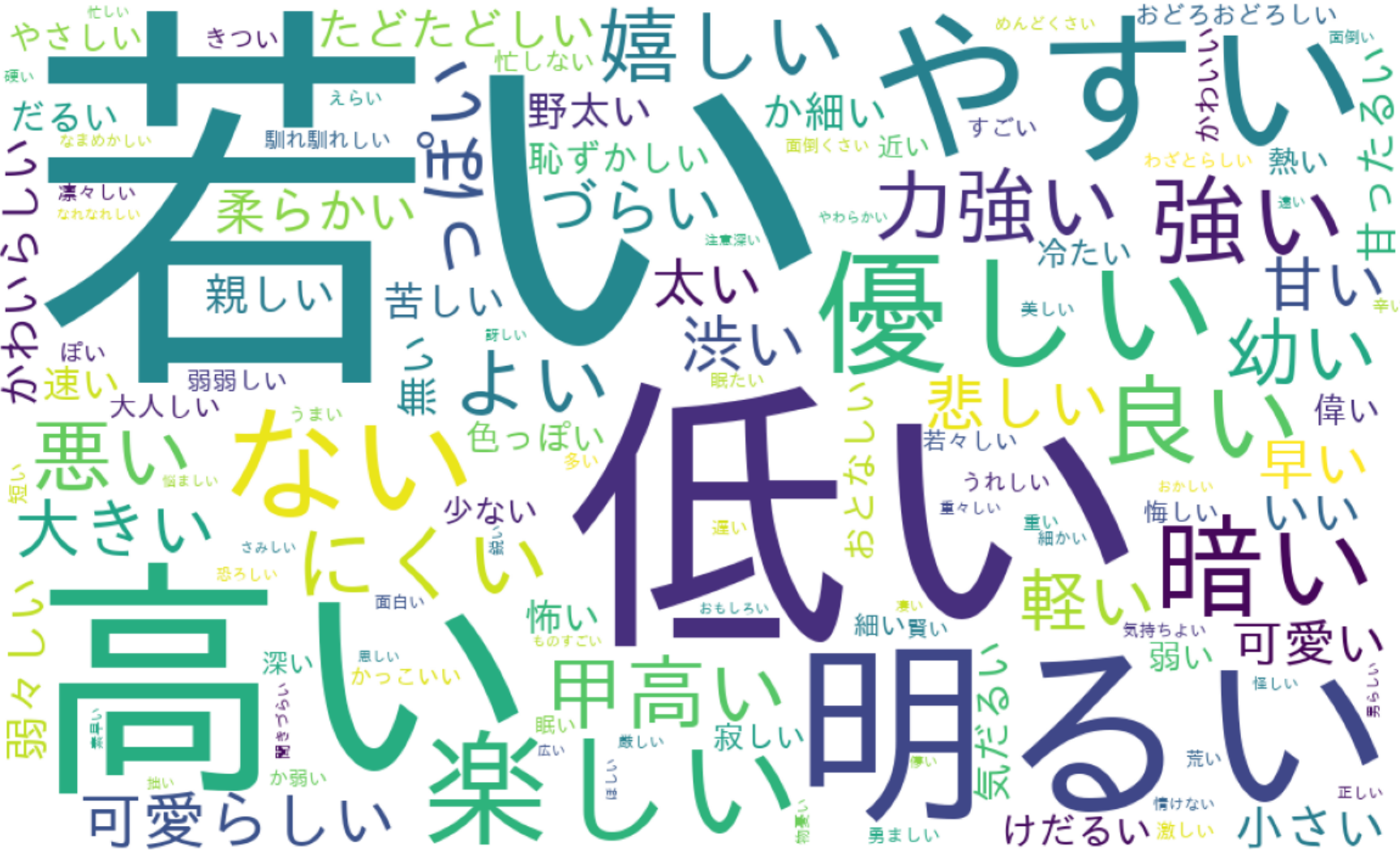}
    \end{minipage}
    \caption{Word cloud of collected voice characteristics descriptions (nouns, verbs, and adjectives, from left).}
    \label{fig:fig/wordcloud.pdf}
\end{figure*}
\begin{table*}[t]
\centering 
\caption{Examples of voice characteristics descriptions.}
\footnotesize
\begin{tabular}{c|p{0.41\linewidth} p{0.4\linewidth}}
    Word & Voice characteristics descriptions & Translations \\
    \hline
    \jp{早口}        & \jp{中年の男性が、ハキハキした声で、早口で喋っている。}& The middle-aged man is speaking rapidly and with a clear voice.\\
    (fast-speaking) & \jp{若い女性が少し慌てた様子で早口で話している。} & A young woman speaks hurriedly with a slightly flustered demeanor.\\
    \hline
    \jp{落ち着く}    & \jp{高齢の女性が低い落ち着いた声でゆっくり喋っている。} & An elderly woman speaks slowly in a calm, low voice.\\
    (to calm down)  & \jp{壮年の男性が、高い声で、落ち着いて説明するように喋っている。} & A middle-aged man speaks with a high-pitched voice, calmly delivering an explanation.\\
    \hline
    \jp{若い} & \jp{若そうな男性が、セリフのようなことをしゃべっている。}& A youthful male speaks in a scripted manner.\\
    (young)     & \jp{若い女性が少年のような口調で喋っている。} & A young woman speaks with a boyish tone.
\end{tabular}
\label{tab:prompt_example}
\end{table*}

\subsubsection{Voice characteristics of video categories}
To explore the relationship between $x$-vectors and video categories, we provide a t-SNE visualization of $x$-vectors colored by video category in \Fig{figure/xvector-category.pdf}. In the Entertainment and Education categories, distinct clusters are noticeable, particularly in the bottom-right and top-central regions. This suggests that typical voice characteristics are aggregated within each category. However, for the majority of the scatter plot, no prominent clusters are observed. This indicates that the speech in this corpus encompasses both typical voices within categories and voices that are shared across categories.

\subsubsection{Linguistic analysis}
We conducted an analysis of word frequency in voice characteristics descriptions tokenized by MeCab. Trivial words such as ``\jp{いる}'' (to be), ``\jp{する}'' (to do), and ``\jp{ある}'' (to exist) were excluded, as well as words that frequently appear due to collection conditions, such as ``\jp{男}''(male) and ``\jp{女}''(female), ``\jp{声}'' (voice), and ``\jp{喋る}'' (to speak). A word cloud visualizing the frequency is presented in \Fig{fig/wordcloud.pdf}. Additionally, examples containing the top-frequency words ``\jp{早口}'' (fast-speaking), ``\jp{落ち着く}'' (to calm down), and ``\jp{若い}'' (young) within each part of speech are provided in \Table{prompt_example}.

High-frequency words not only indicate prevalent voice characteristics features within the corpus but also highlight aspects that are often observed in voice characteristics descriptions. For example, the high frequency of ``\jp{早口}'' as a noun suggests the prevalence of fast-speaking speakers in the YouTube domain and emphasizes the aspect of speech rate, which tends to draw attention.
Furthermore, expressions related to multiple qualitative features also frequently appear. For example, ``\jp{落ち着く}'' is associated with both the depth of the voice, as in ``\jp{落ち着いた低い声}'' (a calm low voice), and with speaking at a calm and slow pace, as in ``\jp{落ち着いてゆっくり喋っている}'' (speaking within a calm, slow voice). However, there are also instances where phrases like ``\jp{高い声で、落ち着いて}'' (calmly with a high voice) exist, indicating the usage of the word ``\jp{落ち着く}'' even with a higher pitch, suggesting the necessity for efforts to predict voice characteristics from expressions that frequently appear across evaluators and are associated with qualitative features but are not entirely deterministic.
\section{Training algorithm of model to retrieve speech from voice characteristics description}
\label{sec:retrieval-model}

We build a model to retrieve speech from voice characteristics descriptions. The straightforward approach would be to implement a contrastive learning such as CLAP, as explained in \Sec{contrastive-learning}. However, since the naive contrastive learning just makes the text embedding and audio embedding closer, the learned embedding is not guaranteed to accurately represent voice characteristics. Therefore, inspired by \cite{yeh2023flap}, we propose an additional training objective: the prediction of speech features from embeddings. The idea here is that the use of perceptible speech features for voice characteristics will enhance the training.

\subsection{Model construction}

\drawfig{t}{0.9\linewidth}{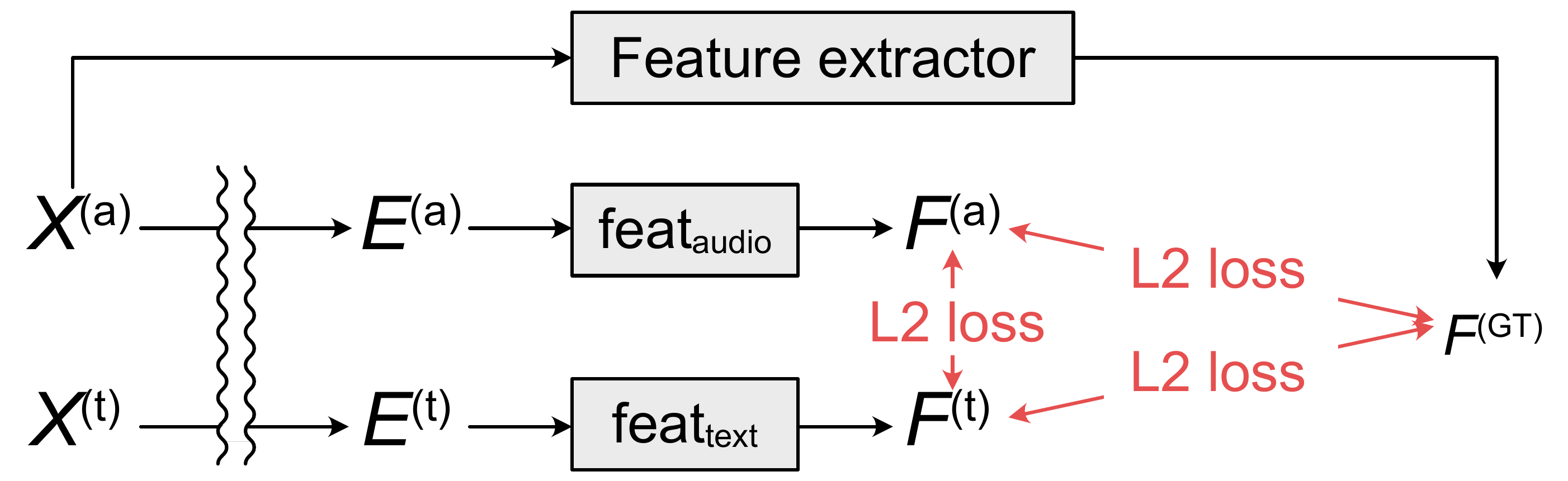}{Feature prediction learning.}
\label{sec:experiments-baseline}
The model architecture adheres to CLAP~\cite{elizalde22clap}. However, while CLAP employs HTS-AT~\cite{chen2022hts}, which is specialized for environmental sounds, as $f_{\text{audio}}(\cdot)$, we use HuBERT~\cite{hsu2021hubert}. Also, we use RoBERTa~\cite{liu2019roberta} as $f_{\text{text}}(\cdot)$.

The basic objective of the model training is contrastive learning, and we add the future prediction learning specialized for voice characteristics. We use networks $ \text{feat}_{\text{audio}}(\cdot)$ and $ \text{feat}_{\text{text}}(\cdot)$, which transform the model outputs $E^{(\text{a})}$ and $E^{(\text{t})}$ into predicted feature vectors $F^{(\text{a})}$ and $F^{(\text{t})}$, respectively. The training criterion is to minimize $L_{\text{feat}}$, where the ground-truth speech features obtained from the $i$th pair data $(X^{(\text{a})}_i, X^{(\text{t})}_i)$ are denoted as $F^{(\text{GT})}_i$, and the predicted feature vectors are $F^{(\text{a})}_i$ and $F^{(\text{t})}_i$.
\begin{align}
    L_{\rm{feat\mathchar`-audio}} &= \sum_{i=1}^N ||F^{(\rm{GT})}_i - F^{(\rm{a})}_i ||_2 \label{eq:feat-audio}\\
    L_{\rm{feat\mathchar`-text}} &= \sum_{i=1}^N ||F^{(\rm{GT})}_i - F^{(\rm{t})}_i ||_2 \label{eq:feat-text}\\
    L_{\rm{feat\mathchar`-cross}} &= \sum_{i=1}^N ||F^{(\rm{a})}_i - F^{(\rm{t})}_i ||_2 \label{eq:feat-cross}\\
    L_{\rm{feat}} &= L_{\rm{feat\mathchar`-audio}} +L_{\rm{feat\mathchar`-text}} +L_{\rm{feat\mathchar`-cross}} \label{eq:feat-loss}
\end{align}
Eqs.\ref{eq:feat-audio} and \ref{eq:feat-text} represent the distances between the predictions from $E^{(\text{a})}$ and $E^{(\text{t})}$ respectively, and the ground-truth features. \Eq{feat-cross} indicates the difference between predictions.
\Fig{figure/feature_extractor.pdf} shows the overview of this learning.

The total training objective is the sum of the contrastive learning loss $L_{\text{CLAP}}$ (Eq. \ref{eq:clap}) and the feature prediction loss $L_{\text{feat}}$ with a coefficient $\alpha$:
\begin{align}
    L = L_{\rm{CLAP}} + \alpha L_{\rm{feat}}.
\end{align}

The gradients of not only $L_{\text{CLAP}}$ but also $L_{\text{feat}}$ are backpropagated. Therefore, $\text{MLP}_{\text{audio}}(\cdot)$ and $\text{MLP}_{\text{feat}}(\cdot)$ are trained to learn embeddings that are relevant to voice characteristics.

\subsection{Evaluation tasks}
\drawfig{t}{0.9\linewidth}{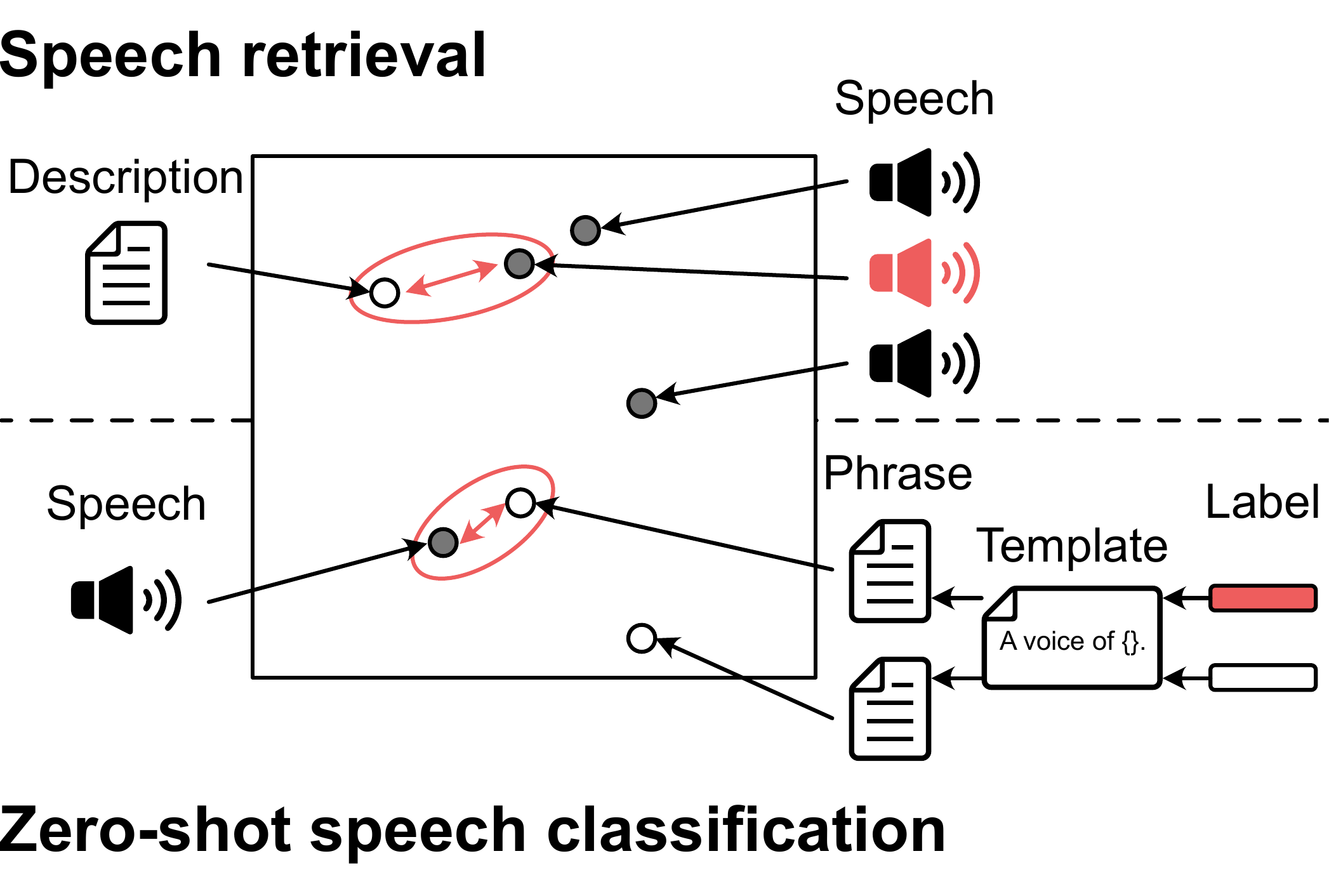}{Overview of evaluation tasks. Top: speech retrieval from voice characteristics description. Bottom: zero-shot speech classification.}

Following the CLAP paper~\cite{elizalde22clap}, we evaluate the trained model and obtained embeddings. \Fig{figure/retrieval.pdf} presents an overview of the evaluation tasks.

\textbf{Speech retrieval from the voice characteristics description.}
We calculate the cosine similarity between the embeddings of the input description and the set of embeddings of target speech segments. A higher cosine similarity value indicates a higher-ranked retrieval result. We evaluate whether the proper segment can be retrieved by the description.

\textbf{Zero-shot speech classification.}
We generate template-based voice characteristics phrases, such as ``a voice of [label].'' Then, a phrase closest to the audio segment in the embedding space is selected.  The label associated with that phrase is considered the classification label for that speech. We evaluate whether the correct label can be obtained without additional training.

\subsection{Experimental setup}
The model was built based on an open-source implementation \footnote{\url{https://github.com/LAION-AI/CLAP}}. We used RoBERTa and HuBERT pre-trained models for Japanese\footnote{\url{https://huggingface.co/rinna/japanese-roberta-base}}\footnote{\url{https://huggingface.co/rinna/japanese-hubert-base}}. The encoder weights were frozen during training. Training/validation/test sets followed the Coco-Nut settings. The embedding dimensionality was set to $512$, and the initial value of the temperature parameter $\tau$ was set to $1/\log(1/0.07)$.
For speech feature prediction, we utilized a $3$-dimensional vector consisting of utterance-level $F_0$ mean, utterance-level standard deviation of energy, and speaking rate. Since our target language is Japanese, a mora-timed language, the speaking rate is calculated as the number of moras per second in an utterance.
These features were chosen based on the analysis in \Sec{linguistic_analysis}, which identified them as being closely associated with perceptually salient voice characteristics.
The feature prediction network consisted of two linear layers with a ReLU activation function~\cite{glorot2011deep} sandwiched in between.
The coefficient $\alpha$ for the feature prediction loss function was set to $0.0$, $0.5$, and $1.0$. We confirmed that the values of $L_{\text{clap}}$ and $L_{\text{feat}}$ are approximately within the same range. Therefore, $\alpha = 1.0$ corresponds to equally weighting these losses. In contrast, $\alpha=0.0$ indicates that $L_{\text{feat}}$ is not considered at all.
The learning rate was set to $5 \times 10^{-6}$, the batch size to $48$, and the number of training epochs to $90$, with model checkpoints saved every $5$ epochs.

Prior to experimental evaluations, a preliminary investigation was conducted to determine $\alpha$. Subsequently, subjective evaluations of speech retrieval and objective evaluations of zero-shot speech classification were performed as the experimental evaluation.

\subsection{Preliminary experiment for $\alpha$}
We conducted a preliminary experiment to investigate the impact of $\alpha$ and determine the optimal number of training epochs.
Gender identification in the top $10$ retrievals was computed at each epoch of model training. The identification score indicates how well the model classified segments of the Coco-Nut validation set to the correct gender. The identification is binary; only male and female gender labels were used.

\drawfig{t}{0.9\linewidth}{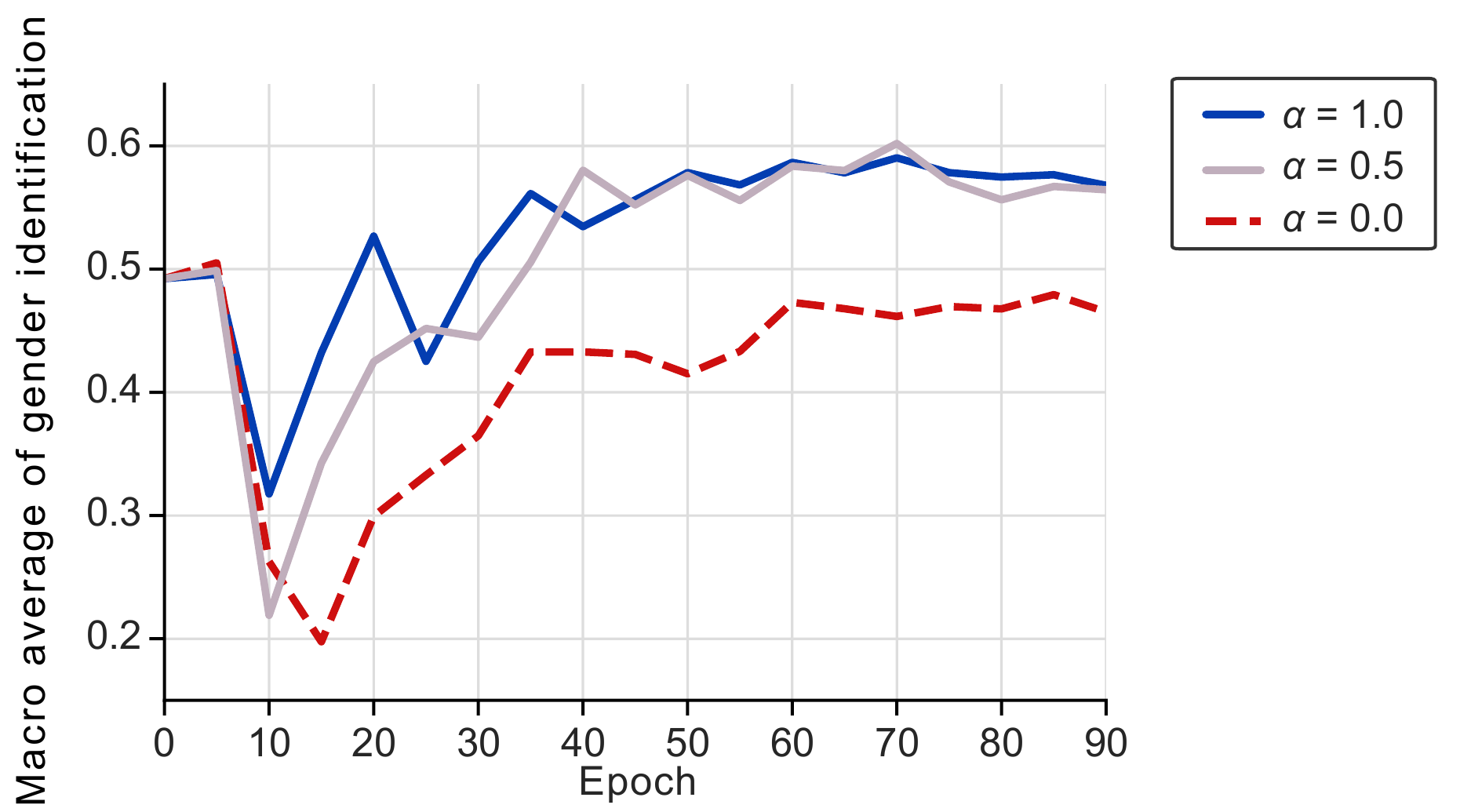}{Macro average gender accuracy among the top $10$ acquired results.}

The results are presented in \Fig{figure/objective_gender.pdf}. 

From the results, the performance ceilings are generally similar for $\alpha = 1.0$ and $\alpha = 0.5$. Moreover, $\alpha > 0$ settings converge faster and better than $\alpha = 0.0$. Hereafter, we used the best settings of model training: $35$ epochs for $\alpha = 1.0$ and $60$ epochs for $\alpha = 0.0$.

\subsection{Subjective evaluation on speech retrieval from the voice characteristics description}
\label{sec:subjective-evaluation}
Subjective evaluation of the speech retrieval results involved subjective evaluation of the correspondence between the voice characteristics description and the retrieved speech segment. The Coco-Nut test set was used for evaluation, utilizing $100$ randomly selected voice characteristics descriptions and targeting all speech segments in the evaluation set for retrieval. The combined voice characteristics description and retrieved speech segment were presented for evaluation of their correspondence on a $9$-point scale ranging from $1$ (no correspondence) to $9$ (excellent correspondence). In addition to the top-ranked retrieval result, experiments were conducted for the $2$nd to $5$th and $6$th to $10$th retrieval results. Moreover, experiments were performed for pairs of description and speech segment selected completely randomly from the evaluation set, as well as for correct or ground-truth pairs within the test set. Listeners were recruited through Lancers\footnote{\url{https://www.lancers.jp/}}, with $16$ evaluations assigned per listener and $20$ listeners assigned per pair.

\drawfig{t}{0.95\linewidth}{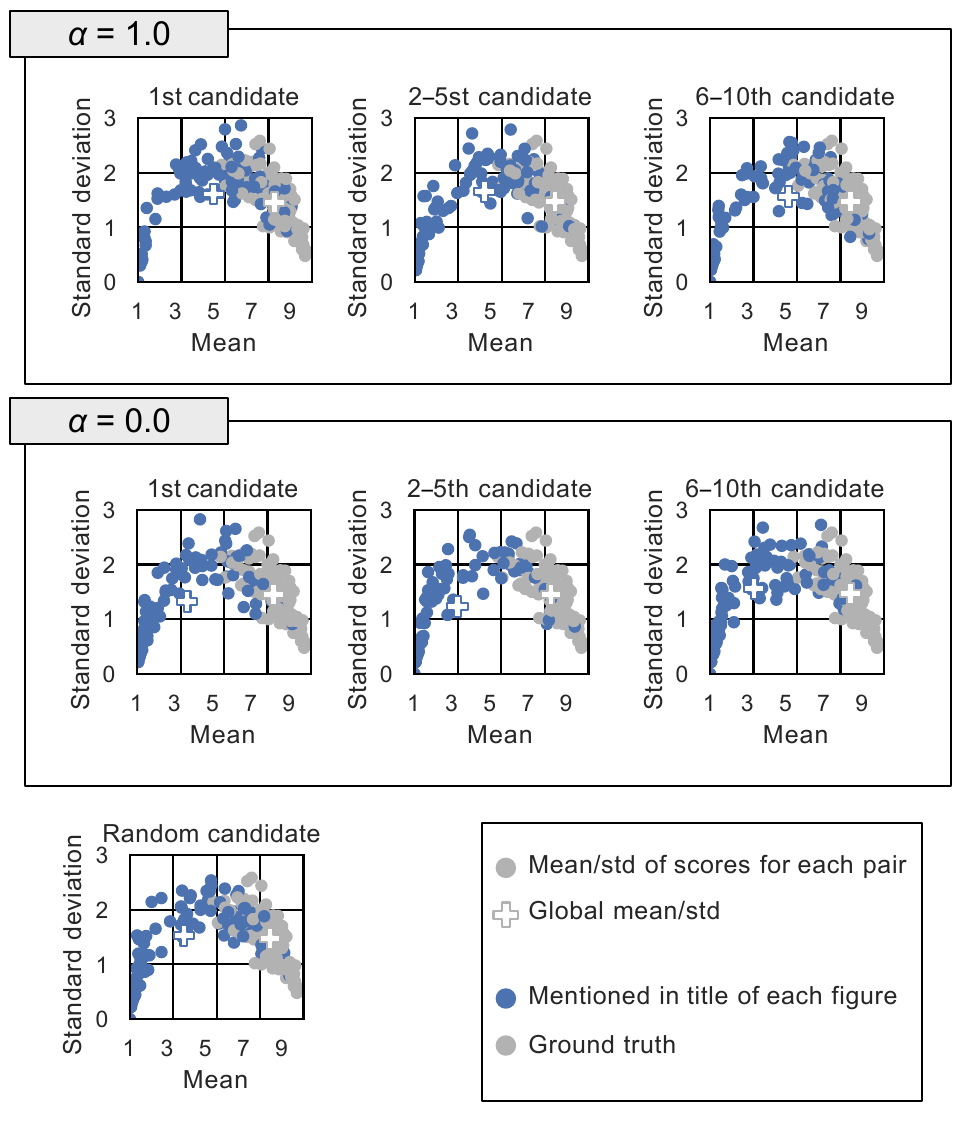}{Mean and standard deviation of subjective evaluation on each description-speech pair. Blue circles indicate retrieved pairs of the figure title, and gray ones indicates ground-truth. ``$+$'' marks indicate average mean and standard deviation of same color plots.}
\drawfig{t}{0.9\linewidth}{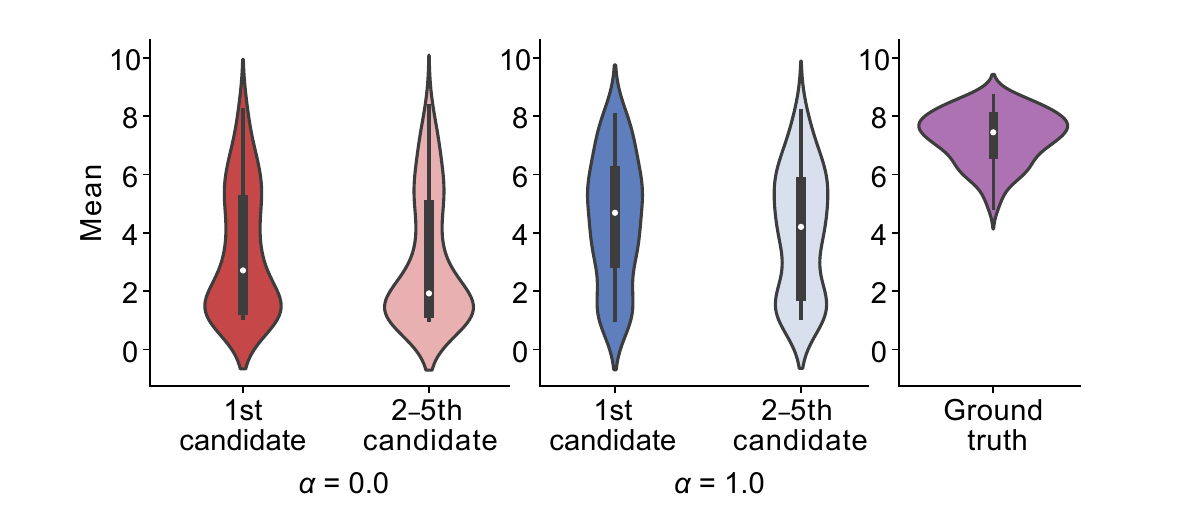}{Subjective evaluation results, showing the distribution of average ratings for each pair of models and retrieval methods.}

\begin{table*}[t]
\centering 
\caption{Examples of retrieval. ``retrieval candidate'' indicates the voice characteristics description paired with retrieved speech.}
\footnotesize
\begin{tabular}{
p{0.2\linewidth} p{0.06\linewidth}|
p{0.24\linewidth} p{0.05\linewidth}
p{0.26\linewidth} p{0.05\linewidth}}
 Retrieval text & Score (GT) & 1st retrieval candidate ($\alpha = 0.0$) & Score & 1st retrieval candidate ($\alpha = 1.0$) & Score \\
 \hline
 \jp{20代くらいの若い女性が楽しそうな声で訴えるようなしゃべり方をしている。} & $6.75$ &
 \jp{20代くらいの女性が、コソコソした声で、指示するように話している。} & $4.86$ &
 \jp{若い女性が、明るくはきはきした声で、少年のように喋っている。} & $5.74$ \\
 (A young woman in her twenties speaks with an appealing tone, expressing herself in an engaging manner.) &&
 (A woman in her twenties speaks in a hushed tone, giving instructions in a clandestine manner.) &&
 (The young woman speaks in a bright and brisk tone, resembling that of a boy.) & \\
 \hline
 \jp{若い男性が、早口で、何かを説明しながら喋っている。} & $8.55$ & 
 \jp{３０代男性が抑えた口調で子供に語りかけている。} & $5.90$ & 
 \jp{50代の男性が、明瞭な声で少し迷いながらも落ち着いて説明している。} & $6.57$ \\
 (The young man speaks rapidly while explaining something.) &&
 (A man in his thirties speaks to a child in a restrained tone.) &&
 (A man in his fifties is speaking clearly with a slightly hesitant yet composed manner while explaining.) & \\
 \hline
 \jp{10代くらいの若い男性が友達と話すような声のトーンで楽しそうに喋っている。} & $7.58$ & 
 \jp{若い女性がのんびりした声で、楽しそうに自分の話に笑いながら喋っている。} & $1.95$ &
 \jp{20代の男性が感情を押さえながら話している。} & $3.19$ \\
 (A young man in his teens speaks with a tone resembling a conversation with friends, expressing excitement.) &&
 (The young woman speaks in a relaxed tone, laughing while joyfully narrating her story.) &&
 (A man in his twenties speaks while controlling his emotions.) &
\end{tabular}
\label{tab:retrieved-examples}
\end{table*}

The results are illustrated in \Fig{figure/subjective.pdf}. It can be observed that, across all rankings from $1$st to $10$th, there is a general trend indicating better scores when using feature prediction learning. This can be attributed to the distribution of average ratings per pair shown in \Fig{figure/subjective_violin.pdf}. It is evident that there are numerous low-rated pairs when feature prediction learning is not used, whereas there is a higher proportion of moderately rated pairs when it is used. This suggests that by incorporating a training objective related to voice characteristics, the occurrence of discordant retrieval results is reduced, leading to a more favorable subjective evaluation.

Examples of retrieval results are provided in \Table{retrieved-examples}. ``Scores (GT)'' refers to the average evaluation score for ground-truth pairs. As seen in the example in the third row, evaluations for voice clips with mismatched gender compared to the voice quality expression sentences tend to be lower, regardless of other matching elements. The improvement in gender match rate as shown in \Fig{figure/objective_gender.pdf} likely contributed to the overall stability of evaluations. In essence, feature prediction learning serves to effectively reflect easily perceptible elements such as gender, thereby contributing to the improvement of performance.

However, even in the case of $1$st results, the scores of ground truth pairs are significantly higher compared to the retrieved results. This is potentially due to the high quality of the collected voice characteristics descriptions, but it also suggests room for improvement in the proposed method.

\subsection{Objective evaluation on zero-shot speech classification}
Objective evaluation involved assessing binary label classification performance using gender and speaking rate. The evaluation targeted the parallel100 set from JVS~\cite{takamichi2020jsut}, comprising a gender label and speaking rate of each of $100$ speakers. The phrases were ``\jp{男性が喋っている}'' (a man is speaking) and ``\jp{女性が喋っている}'' (a woman is speaking) for gender classification and ``\jp{ゆっくり喋っている}'' (speaking slowly) and ``\jp{早口で喋っている}'' (speaking rapidly) for speaking rate.
We used all $100$ speakers for gender identification. For speaking rate identification, we used the $20$ fastest-speaking and $20$ slowest-speaking speakers.
Since the JVS fast-speaking speakers speak slower than Coco-Nut fast-speaking ones, we sped up the speech of fast-speaking speakers to $1.5$ times their original speed using WSOLA~\cite{verhelst1993overlap}.

\begin{table}[t]
    \centering
    \caption{Results of zero-shot classification.}
        \begin{tabularx}{0.6\linewidth}{c|c|c}
            Label & $\alpha$ & Accuracy \\
            \hline
            \multirow{2}{*}{Male / Female}  & $0.0$ & $0.998$ / $0.941$ \\
                                            & $1.0$ & $0.986$ / $1.000$ \\ \hline
            \multirow{2}{*}{Fast / Slow}    & $0.0$ & $0.984$ / $0.997$ \\
                                            & $1.0$ & $0.990$ / $1.000$
        \end{tabularx}
    \label{tab:zero-shot_gender_speaking-rate}
\end{table}

The results are shown in \Table{zero-shot_gender_speaking-rate}. Here, utilizing feature prediction learning results in higher accuracy rates overall.

\section{Conclusion} \label{sec:conclusion}
In this paper, we developed a paired corpus of speech and voice characteristics descriptions. This corpus is now available publicly and will promote research on prompt-based TTS. Also, we proposed a retrieval model training incorporating feature prediction learning into contrastive learning and experimentally validated its performance.
We leave the construction of prompt-based TTS with the proposed corpus and retrieval model to future work.


%

\appendices

\section*{Acknowledgment}
This work was supported by JSPS KAKENHI 21H04900, 22H03639, 23H03418 (practical experiment), JST FOREST JPMJFR226V, and Moon-shot R\&D Grant Number JPMJPS2011 (algorithm development). We appreciate Takaaki Saeki and Yuta Matsunaga of the University of Tokyo for their help.
We also wish to thank Masaki Sato, Kazuki Yamauchi, Hiroaki Hyodo, Osamu Take of the University of Tokyo for their assistance in manual quality assurance.

\ifCLASSOPTIONcaptionsoff
  \newpage
\fi



\bibliographystyle{IEEEtran}
%
\bibliography{bibtex/refs}

\begin{thebibliography}{10}
\providecommand{\url}[1]{#1}
\csname url@samestyle\endcsname
\providecommand{\newblock}{\relax}
\providecommand{\bibinfo}[2]{#2}
\providecommand{\BIBentrySTDinterwordspacing}{\spaceskip=0pt\relax}
\providecommand{\BIBentryALTinterwordstretchfactor}{4}
\providecommand{\BIBentryALTinterwordspacing}{\spaceskip=\fontdimen2\font plus
\BIBentryALTinterwordstretchfactor\fontdimen3\font minus \fontdimen4\font\relax}
\providecommand{\BIBforeignlanguage}[2]{{%
\expandafter\ifx\csname l@#1\endcsname\relax
\typeout{** WARNING: IEEEtran.bst: No hyphenation pattern has been}%
\typeout{** loaded for the language `#1'. Using the pattern for}%
\typeout{** the default language instead.}%
\else
\language=\csname l@#1\endcsname
\fi
#2}}
\providecommand{\BIBdecl}{\relax}
\BIBdecl

\bibitem{oord2016wavenet}
A.~{van den Oord}, S.~Dieleman, H.~Zen, K.~Simonyan, O.~Vinyals, A.~Graves, N.~Kalchbrenner, A.~Senior, and K.~Kavukcuoglu, ``{WaveNet}: A generative model for raw audio,'' in \emph{{Proc. SSW}}, Sunnyvale, U.S.A., 2016, p. 125.

\bibitem{wang2017tacotron}
Y.~Wang, R.~Skerry-Ryan, D.~Stanton, Y.~Wu, R.~J. Weiss, N.~Jaitly, Z.~Yang, Y.~Xiao, Z.~Chen, S.~Bengio, Q.~Le, Y.~Agiomyrgiannakis, R.~Clark, and R.~A. Saurous, ``{Tacotron: Towards End-to-End Speech Synthesis},'' in \emph{{Proc. INTERSPEECH}}, Stockholm, Sweden, 2017, pp. 4006--4010.

\bibitem{ren2019fastspeech}
Y.~Ren, Y.~Ruan, X.~Tan, T.~Qin, S.~Zhao, Z.~Zhao, and T.-Y. Liu, ``{FastSpeech}: Fast, robust and controllable text to speech,'' \emph{{Proc. NeurIPS}}, vol.~32, 2019.

\bibitem{hojo16speakercode}
N.~Hojo, Y.~Ijima, and H.~Mizuno, ``{DNN}-based speech synthesis using speaker codes,'' \emph{IEICE TRANSACTIONS on Information and Systems}, vol. E101-D, no.~2, pp. 462--472, 2017.

\bibitem{stanton2022speaker}
D.~Stanton, M.~Shannon, S.~Mariooryad, R.~Skerry-Ryan, E.~Battenberg, T.~Bagby, and D.~Kao, ``Speaker generation,'' in \emph{{Proc. ICASSP}}, 2022, pp. 7897--7901.

\bibitem{watanabe22mid-attribute-speaker-generation}
A.~Watanabe, S.~Takamichi, Y.~Saito, D.~Xin, and H.~Saruwatari, ``Mid-attribute speaker generation using optimal-transport-based interpolation of gaussian mixture models,'' in \emph{{Proc. ICASSP}}.\hskip 1em plus 0.5em minus 0.4em\relax IEEE, 2023, pp. 1--5.

\bibitem{gustafson21personality-in-the-mix}
J.~Gustafson, J.~Beskow, and E.~Szekely, ``{Personality in the mix - investigating the contribution of fillers and speaking style to the perception of spontaneous speech synthesis},'' in \emph{Proc. Speech Synthesis Workshop}, 2021, pp. 48--53.

\bibitem{zhang19learning-to-speak}
Y.~Zhang, R.~J. Weiss, H.~Zen, Y.~Wu, Z.~Chen, R.~Skerry-Ryan, Y.~Jia, A.~Rosenberg, and B.~Ramabhadran, ``{Learning to Speak Fluently in a Foreign Language: Multilingual Speech Synthesis and Cross-Language Voice Cloning},'' in \emph{Proc. Interspeech}, Graz, Austria, Sep. 2019, pp. 2080--2084.

\bibitem{rui21reinforcement-learning-emotional-tts}
R.~Liu, B.~Sisman, and H.~Li, ``Reinforcement learning for emotional text-to-speech synthesis with improved emotion discriminability,'' in \emph{Proc. Interspeech}, Aug. 2021, pp. 4648--4652.

\bibitem{ohta10voice-quality-gmm-vc}
K.~Ohta, T.~Toda, Y.~Ohtani, H.~Saruwatari, and K.~Shikano, ``{Adaptive voice-quality control based on one-to-many eigenvoice conversion},'' in \emph{Proc. Interspeech}, 2010, pp. 2158--2161.

\bibitem{raitio20controllable-tts}
T.~Raitio, R.~Rasipuram, and D.~Castellani, ``{Controllable Neural Text-to-Speech Synthesis Using Intuitive Prosodic Features},'' in \emph{Proc. Interspeech}, 2020, pp. 4432--4436.

\bibitem{ramesh21dalle}
A.~Ramesh, M.~Pavlov, G.~Goh, S.~Gray, C.~Voss, A.~Radford, M.~Chen, and I.~Sutskever, ``Zero-shot text-to-image generation,'' \emph{arXiv:2102.12092}, 2021.

\bibitem{elizalde22clap}
B.~Elizalde, S.~Deshmukh, M.~A. Ismail, and H.~Wang, ``{CLAP}: Learning audio concepts from natural language supervision,'' \emph{arXiv:2206.04769}, 2022.

\bibitem{huang22mulan}
Q.~Huang, A.~Jansen, J.~Lee, R.~Ganti, J.~Y. Li, and D.~P.~W. Ellis, ``{MuLan}: A joint embedding of music audio and natural language,'' \emph{arXiv:2208.12415}, 2022.

\bibitem{ho22imagenvideo}
J.~Ho, W.~Chan, C.~Saharia, J.~Whang, R.~Gao, A.~Gritsenko, D.~P. Kingma, B.~Poole, M.~Norouzi, D.~J. Fleet, and T.~Salimans, ``{Imagen Video}: High definition video generation with diffusion models,'' \emph{arXiv:2210.02303}, 2022.

\bibitem{radford21clip}
\BIBentryALTinterwordspacing
A.~Radford, J.~W. Kim, C.~Hallacy, A.~Ramesh, G.~Goh, S.~Agarwal, G.~Sastry, A.~Askell, P.~Mishkin, J.~Clark, G.~Krueger, and I.~Sutskever, ``Learning transferable visual models from natural language supervision,'' in \emph{Proceedings of the 38th International Conference on Machine Learning}, ser. Proceedings of Machine Learning Research, M.~Meila and T.~Zhang, Eds., vol. 139.\hskip 1em plus 0.5em minus 0.4em\relax PMLR, 18--24 Jul 2021, pp. 8748--8763. [Online]. Available: \url{https://proceedings.mlr.press/v139/radford21a.html}
\BIBentrySTDinterwordspacing

\bibitem{lin14ms-coco}
T.-Y. Lin, M.~Maire, S.~Belongie, L.~Bourdev, R.~Girshick, J.~Hays, P.~Perona, D.~Ramanan, C.~L. Zitnick, and P.~Dollár, ``{Microsoft COCO}: Common objects in context,'' \emph{arXiv:1405.0312}, 2014.

\bibitem{sharma18conceptualcaptions}
\BIBentryALTinterwordspacing
P.~Sharma, N.~Ding, S.~Goodman, and R.~Soricut, ``Conceptual captions: A cleaned, hypernymed, image alt-text dataset for automatic image captioning,'' in \emph{Proceedings of the 56th Annual Meeting of the Association for Computational Linguistics (Volume 1: Long Papers)}.\hskip 1em plus 0.5em minus 0.4em\relax Melbourne, Australia: Association for Computational Linguistics, Jul. 2018, pp. 2556--2565. [Online]. Available: \url{https://aclanthology.org/P18-1238}
\BIBentrySTDinterwordspacing

\bibitem{kim19audiocaps}
C.~D. Kim, B.~Kim, H.~Lee, and G.~Kim, ``Audiocaps: Generating captions for audios in the wild,'' in \emph{NAACL-HLT}, 2019.

\bibitem{drossos20clotho}
K.~Drossos, S.~Lipping, and T.~Virtanen, ``Clotho: an audio captioning dataset,'' in \emph{{Proc. ICASSP}}, 2020, pp. 736--740.

\bibitem{wu22laion-audio}
Y.~Wu, K.~Chen, T.~Zhang, Y.~Hui, T.~Berg-Kirkpatrick, and S.~Dubnov, ``Large-scale contrastive language-audio pretraining with feature fusion and keyword-to-caption augmentation,'' \emph{arXiv:2211.06687}, 2022.

\bibitem{guo22prompttts}
Z.~Guo, Y.~Leng, Y.~Wu, S.~Zhao, and X.~Tan, ``{PromptTTS}: Controllable text-to-speech with text descriptions,'' \emph{arXiv:2211.12171}, 2022.

\bibitem{yang23instructtts}
D.~Yang, S.~Liu, R.~Huang, G.~Lei, C.~Weng, H.~Meng, and D.~Yu, ``{InstructTTS}: Modelling expressive {TTS} in discrete latent space with natural language style prompt,'' \emph{arXiv:2301.13662}, 2023.

\bibitem{zhang2023promptspeaker}
Y.~Zhang, G.~Liu, Y.~Lei, Y.~Chen, H.~Yin, L.~Xie, and Z.~Li, ``Promptspeaker: Speaker generation based on text descriptions,'' \emph{arXiv preprint arXiv:2310.05001}, 2023.

\bibitem{lyth2024natural}
D.~Lyth and S.~King, ``Natural language guidance of high-fidelity text-to-speech with synthetic annotations,'' \emph{arXiv preprint arXiv:2402.01912}, 2024.

\bibitem{zen19libritts}
H.~Zen, V.~Dang, R.~Clark, Y.~Zhang, R.~J. Weiss, Y.~Jia, Z.~Chen, and Y.~Wu, ``{LibriTTS}: A corpus derived from {LibriSpeech} for text-to-speech,'' in \emph{Proc. Interspeech}, 2019, pp. 1526--1530.

\bibitem{takamichi2020jsut}
S.~Takamichi, R.~Sonobe, K.~Mitsui, Y.~Saito, T.~Koriyama, N.~Tanji, and H.~Saruwatari, ``{JSUT and JVS}: Free {Japanese} voice corpora for accelerating speech synthesis research,'' \emph{Acoustical Science and Technology}, vol.~41, no.~5, pp. 761--768, 2020.

\bibitem{watanabe2023coconut}
A.~Watanabe, S.~Takamichi, Y.~Saito, W.~Nakata, D.~Xin, and H.~Saruwatari, ``Coco-nut: Corpus of japanese utterance and voice characteristics description for prompt-based control,'' in \emph{2023 IEEE Automatic Speech Recognition and Understanding Workshop (ASRU)}, 2023, pp. 1--8.

\bibitem{agostinelli23musiclm}
A.~Agostinelli, T.~I. Denk, Z.~Borsos, J.~Engel, M.~Verzetti, A.~Caillon, Q.~Huang, A.~Jansen, A.~Roberts, M.~Tagliasacchi, M.~Sharifi, N.~Zeghidour, and C.~Frank, ``{MusicLM}: Generating music from text,'' \emph{arXiv:2301.11325}, 2023.

\bibitem{ohnaka22visual-onoma-to-wave}
H.~Ohnaka, S.~Takamichi, K.~Imoto, Y.~Okamoto, K.~Fujii, and H.~Saruwatari, ``Visual onoma-to-wave: environmental sound synthesis from visual onomatopoeias and sound-source images,'' in \emph{{Proc. ICASSP}}.\hskip 1em plus 0.5em minus 0.4em\relax IEEE, 2023, pp. 1--5.

\bibitem{laion-ai-audio}
Y.~Wu, K.~Chen, T.~Zhang, M.~Nezhurina, and Y.~Hui, ``{LAION}-{A}udio-630{K},'' 2022, \url{https://github.com/LAION-AI/audio-dataset}.

\bibitem{molad23dreamix}
E.~Molad, E.~Horwitz, D.~Valevski, A.~R. Acha, Y.~Matias, Y.~Pritch, Y.~Leviathan, and Y.~Hoshen, ``Dreamix: Video diffusion models are general video editors,'' \emph{arXiv:2302.01329}, 2023.

\bibitem{nagrani2020voxceleb}
A.~Nagrani, J.~S. Chung, W.~Xie, and A.~Zisserman, ``Voxceleb: Large-scale speaker verification in the wild,'' \emph{Computer Speech \& Language}, vol.~60, p. 101027, 2020.

\bibitem{maekawa2003csj}
K.~Maekawa, ``Corpus of spontaneous japanese : its design and evaluation,'' \emph{Proceedings of The ISCA \& IEEE Workshop on Spontaneous Speech Processing and Recognition (SSPR 2003)}, pp. 7--12, 2003.

\bibitem{yeh2023flap}
C.-F. Yeh, P.-Y. Huang, V.~Sharma, S.-W. Li, and G.~Gosh, ``Flap: Fast language-audio pre-training,'' \emph{arXiv preprint arXiv:2311.01615}, 2023.

\bibitem{chen2021gigaspeech}
G.~Chen, S.~Chai, G.~Wang, J.~Du, W.-Q. Zhang, C.~Weng, D.~Su, D.~Povey, J.~Trmal, J.~Zhang \emph{et~al.}, ``{GigaSpeech}: An evolving, multi-domain asr corpus with 10,000 hours of transcribed audio,'' \emph{arXiv preprint arXiv:2106.06909}, 2021.

\bibitem{takamichi21jtubespeech}
S.~Takamichi, L.~K{\"u}rzinger, T.~Saeki, S.~Shiota, and S.~Watanabe, ``{JTubeSpeech}: corpus of {Japanese} speech collected from {YouTube} for speech recognition and speaker verification,'' \emph{arXiv:2112.09323}, 2021.

\bibitem{yin23reazonspeech}
S.~F. Yue~Yin, Daijiro~Mori, ``{ReazonSpeech}: A free and massive corpus for {J}apanese {ASR},'' in \emph{Annual meetings of the Association for Natural Language Processing}, 2023.

\bibitem{devlin2018bert}
J.~Devlin, M.-W. Chang, K.~Lee, and K.~Toutanova, ``Bert: Pre-training of deep bidirectional transformers for language understanding,'' \emph{arXiv preprint arXiv:1810.04805}, 2018.

\bibitem{ddoukhanicassp2018}
D.~Doukhan, J.~Carrive, F.~Vallet, A.~Larcher, and S.~Meignier, ``An open-source speaker gender detection framework for monitoring gender equality,'' in \emph{{Proc. ICASSP}}.\hskip 1em plus 0.5em minus 0.4em\relax IEEE, 2018.

\bibitem{mittag21nisqa}
G.~Mittag, B.~Naderi, A.~Chehadi, and S.~Möller, ``{NISQA}: A deep {CNN}-self-attention model for multidimensional speech quality prediction with crowdsourced datasets,'' in \emph{Proc. Interspeech}, 2021, pp. 2127--2131.

\bibitem{ward63clustering}
J.~H.~W. Jr., ``Hierarchical grouping to optimize an objective function,'' \emph{Journal of the American Statistical Association}, vol.~58, no. 301, pp. 236--244, 1963.

\bibitem{snyder2018x}
D.~Snyder, D.~Garcia-Romero, G.~Sell, D.~Povey, and S.~Khudanpur, ``X-vectors: Robust dnn embeddings for speaker recognition,'' in \emph{{Proc. ICASSP}}.\hskip 1em plus 0.5em minus 0.4em\relax IEEE, 2018, pp. 5329--5333.

\bibitem{brown2021playing}
A.~Brown, J.~Huh, A.~Nagrani, J.~S. Chung, and A.~Zisserman, ``Playing a part: Speaker verification at the movies,'' in \emph{{Proc. ICASSP}}.\hskip 1em plus 0.5em minus 0.4em\relax IEEE, 2021, pp. 6174--6178.

\bibitem{radford2022robust}
A.~Radford, J.~W. Kim, T.~Xu, G.~Brockman, C.~McLeavey, and I.~Sutskever, ``Robust speech recognition via large-scale weak supervision,'' \emph{arXiv preprint arXiv:2212.04356}, 2022.

\bibitem{salazar20maskedlaugagemodel}
\BIBentryALTinterwordspacing
J.~Salazar, D.~Liang, T.~Q. Nguyen, and K.~Kirchhoff, ``Masked language model scoring,'' in \emph{Proceedings of the 58th Annual Meeting of the Association for Computational Linguistics}.\hskip 1em plus 0.5em minus 0.4em\relax Online: Association for Computational Linguistics, Jul. 2020, pp. 2699--2712. [Online]. Available: \url{https://aclanthology.org/2020.acl-main.240}
\BIBentrySTDinterwordspacing

\bibitem{audiocaps}
C.~D. Kim, B.~Kim, H.~Lee, and G.~Kim, ``Audiocaps: Generating captions for audios in the wild,'' in \emph{NAACL-HLT}, 2019.

\bibitem{chen2022hts}
K.~Chen, X.~Du, B.~Zhu, Z.~Ma, T.~Berg-Kirkpatrick, and S.~Dubnov, ``Hts-at: A hierarchical token-semantic audio transformer for sound classification and detection,'' in \emph{{Proc. ICASSP}}.\hskip 1em plus 0.5em minus 0.4em\relax IEEE, 2022, pp. 646--650.

\bibitem{hsu2021hubert}
W.-N. Hsu, B.~Bolte, Y.-H.~H. Tsai, K.~Lakhotia, R.~Salakhutdinov, and A.~Mohamed, ``Hubert: Self-supervised speech representation learning by masked prediction of hidden units,'' \emph{IEEE/ACM Transactions on Audio, Speech, and Language Processing}, vol.~29, pp. 3451--3460, 2021.

\bibitem{liu2019roberta}
Y.~Liu, M.~Ott, N.~Goyal, J.~Du, M.~Joshi, D.~Chen, O.~Levy, M.~Lewis, L.~Zettlemoyer, and V.~Stoyanov, ``Roberta: A robustly optimized bert pretraining approach,'' \emph{arXiv preprint arXiv:1907.11692}, 2019.

\bibitem{glorot2011deep}
X.~Glorot, A.~Bordes, and Y.~Bengio, ``Deep sparse rectifier neural networks,'' in \emph{Proceedings of the fourteenth international conference on artificial intelligence and statistics}.\hskip 1em plus 0.5em minus 0.4em\relax JMLR Workshop and Conference Proceedings, 2011, pp. 315--323.

\bibitem{verhelst1993overlap}
W.~Verhelst and M.~Roelands, ``An overlap-add technique based on waveform similarity (wsola) for high quality time-scale modification of speech,'' in \emph{1993 IEEE International Conference on Acoustics, Speech, and Signal Processing}, vol.~2.\hskip 1em plus 0.5em minus 0.4em\relax IEEE, 1993, pp. 554--557.

\end{thebibliography}

\begin{IEEEbiography}[{\includegraphics[width=1in,height=1.25in,clip,keepaspectratio]{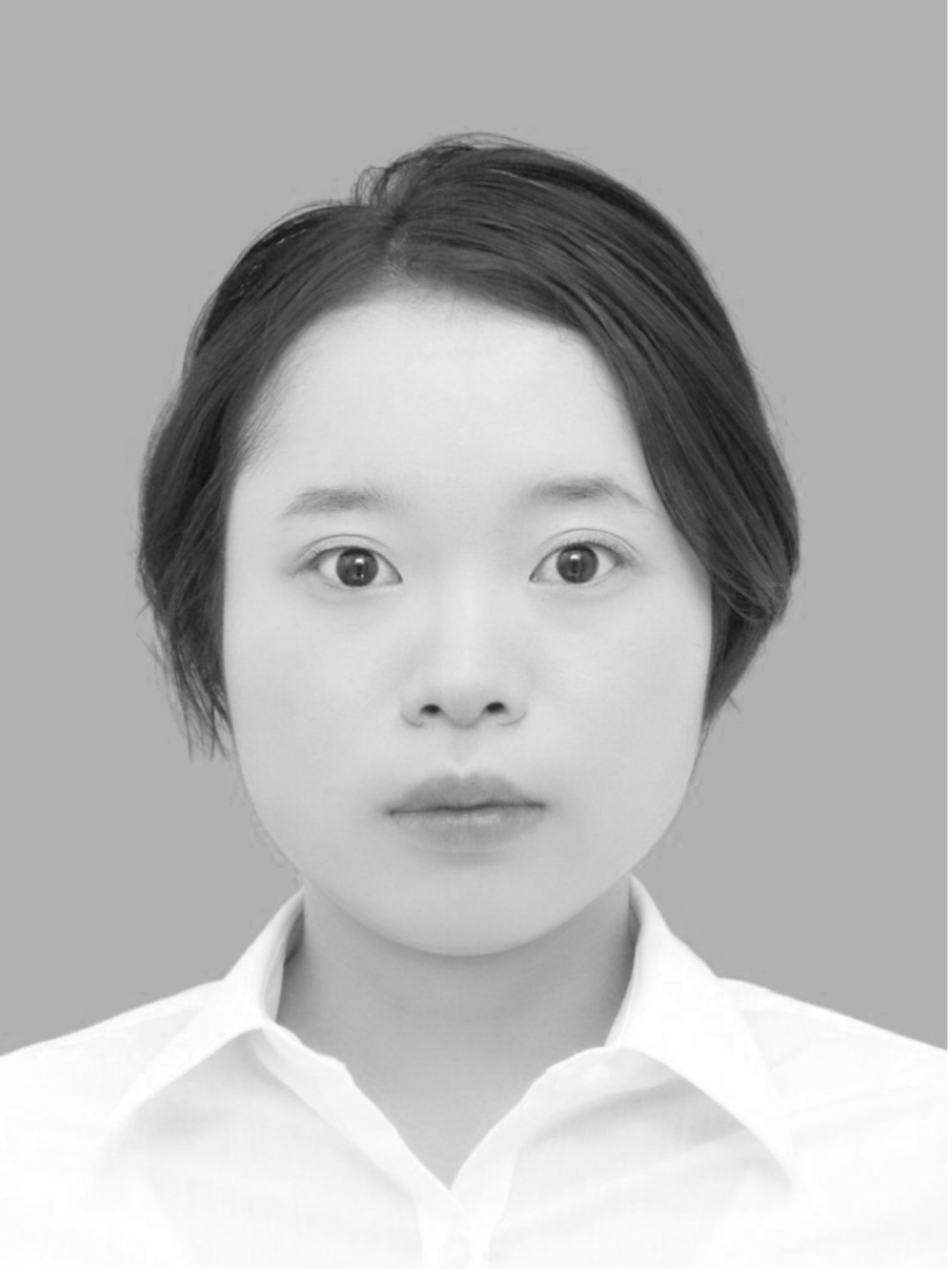}}]{Aya Watanabe}
received a B.E. degree from Waseda University, Tokyo, Japan in 2022. She is now a master's course student at the Graduate School of Information Science and Technology, The University of Tokyo, Tokyo, Japan. Her research interest lies in speech synthesis, with a particular focus on the voice characteristics of synthesized speech. She is currently a Student Member of the Acoustical Society of Japan (ASJ). She received the 27th best student presentation award of ASJ in 2023.
\end{IEEEbiography}

\begin{IEEEbiography}[{\includegraphics[width=1in,height=1.25in,clip,keepaspectratio]{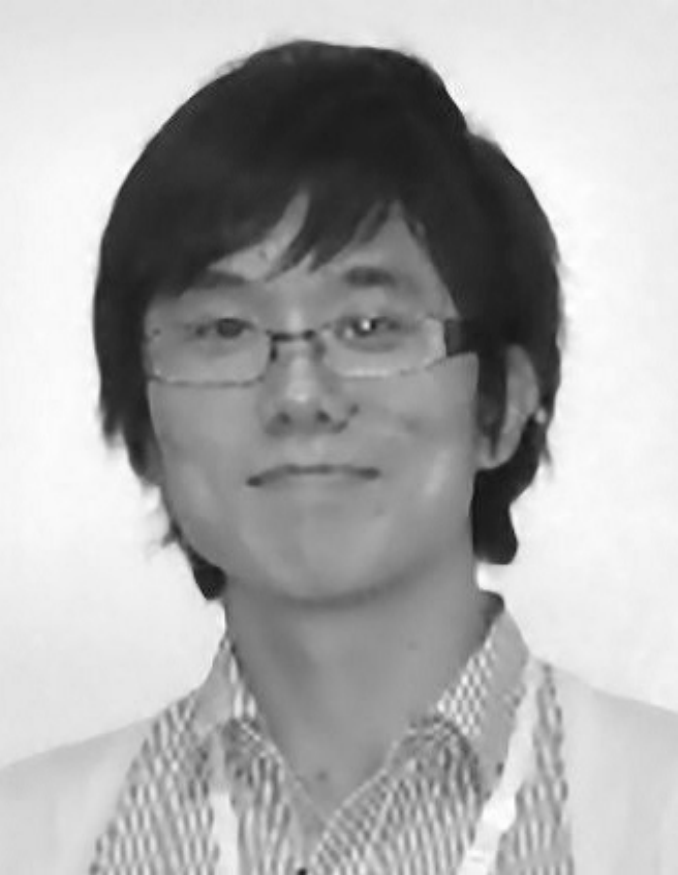}}]{Shinnosuke Takamichi} received the Ph.D. degree from the Graduate School of Information Science, Nara Institute of Science and Technology, Japan, in 2016. He is currently a Lecturer at The University of Tokyo. He has received more than 20 paper/achievement awards including the 2020 IEEE Signal Processing Society Young Author Best Paper Award. 
\end{IEEEbiography}

\begin{IEEEbiography}[{\includegraphics[width=1in,height=1.25in,clip,keepaspectratio]{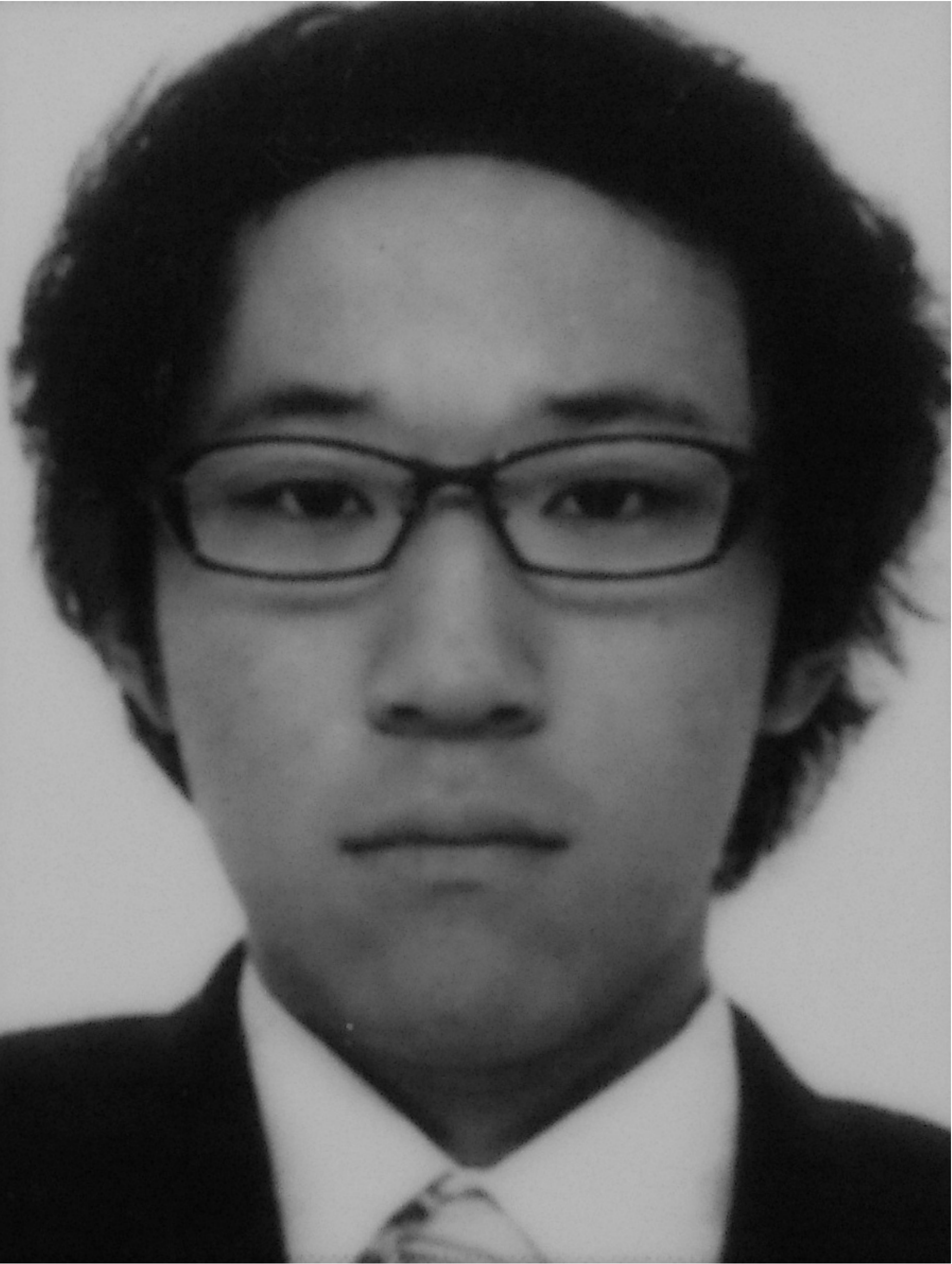}}]{Yuki Saito} received the Ph.D. degree in Information Science and Technology in 2021 from the Graduate School of Information Science and Technology, The University of Tokyo, Japan. His research interests include speech synthesis, voice conversion, and machine learning. He was the recipient of eight paper awards including the 2020 IEEE SPS Young Author Best Paper Award. He is a Member of the Acoustical Society of Japan, a Member of IEEE SPS, and a Member of  Institute of Electronics, Information and Communication Engineers.
\end{IEEEbiography}

\begin{IEEEbiography}
[{\includegraphics[width=1in,height=1.25in,clip,keepaspectratio]{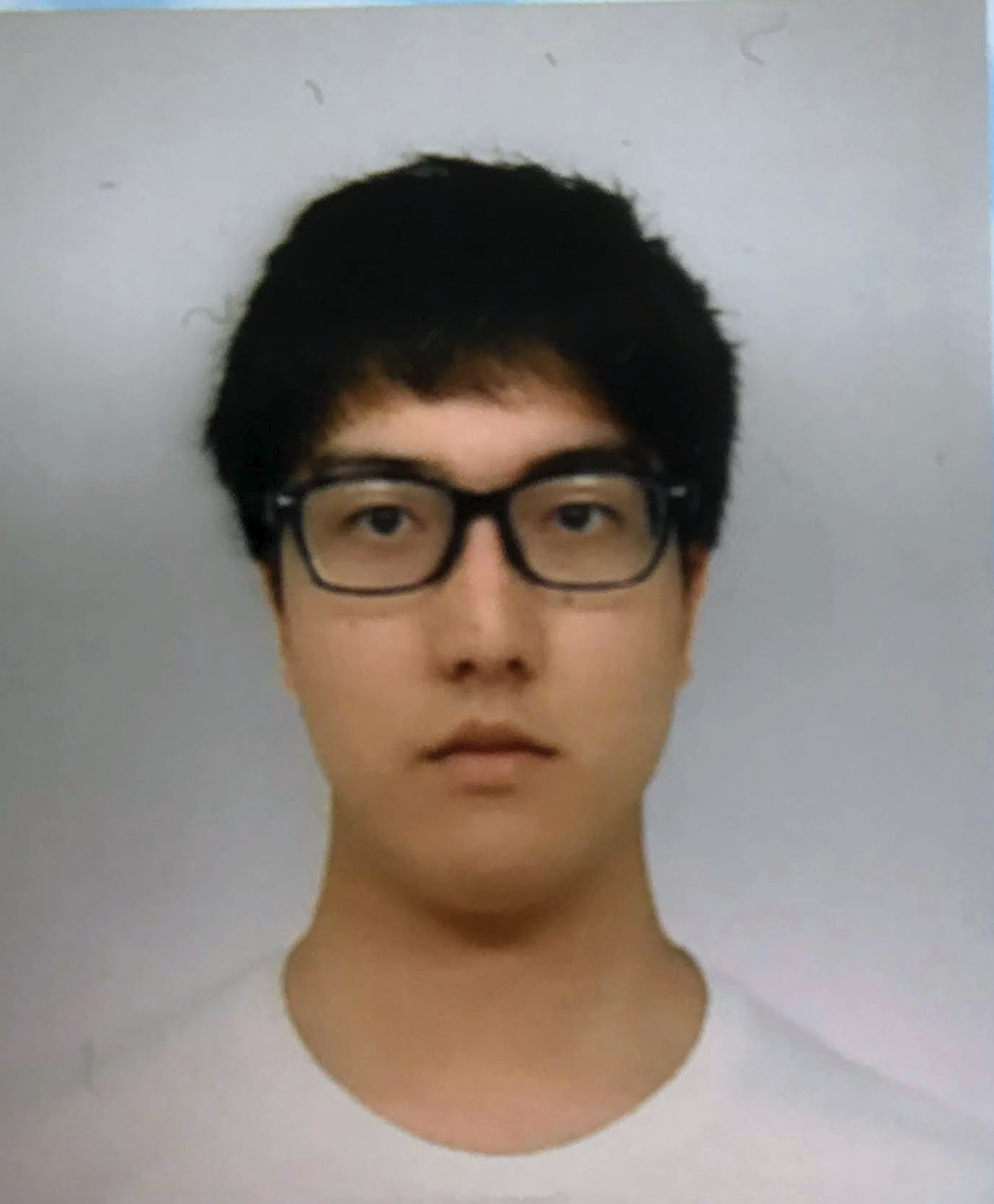}}]{Wataru Nakata} received the B.E. degree from The University of Tokyo, Tokyo, Japan, in 2023. 
He is now a M.E. sutdent at the University of Tokyo.
His research interests include, speech synthesis, natural language processing and deep learning.
He is a Student Member of the Acoustical Society of Japan. 
He recieved the best student presentation award of ASJ in 2022.
\end{IEEEbiography}

\begin{IEEEbiography}
[{\includegraphics[width=1in,height=1.25in,clip,keepaspectratio]{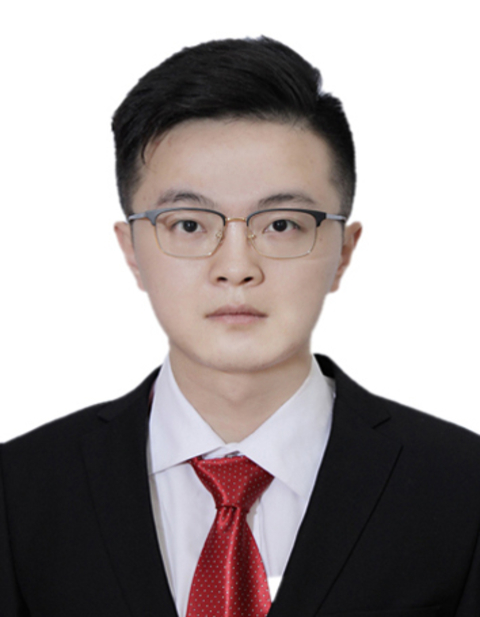}}]
{Detai Xin} received the B.E. degree from Beihang University (former Beijing University of Aeronautics and Astronautics), Beijing, China, in 2019, and the M.E. degree from the Graduate School of Information Science and Technology, The University of Tokyo, Japan, in 2021.
He is now a PhD student at the University of Tokyo.
His research interests include speech synthesis, speech processing, and deep learning.
He has publicated more than 10 papers on speech synthesis and speech processing.
He received the IEEE SPS Tokyo Joint Chapter Student Award in 2022.
\end{IEEEbiography}

\begin{IEEEbiography}[{\includegraphics[width=1in,height=1.25in,clip,keepaspectratio]{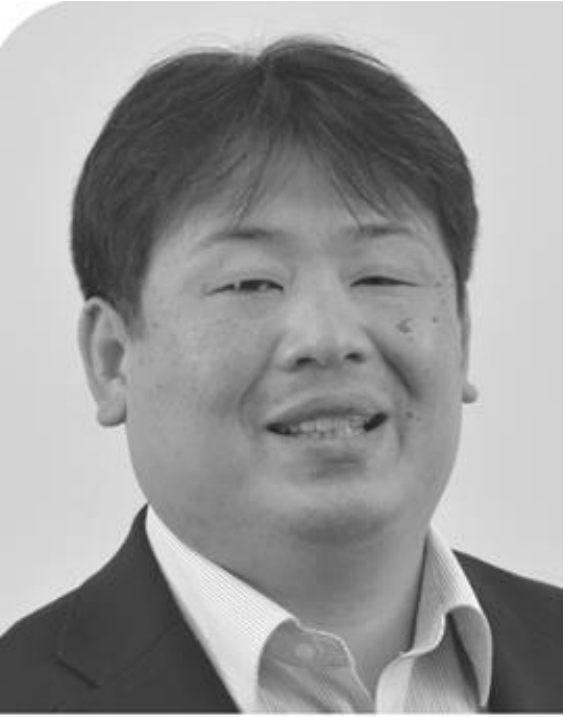}}]{Hiroshi Saruwatari} received the B.E., M.E., and Ph.D. degrees from Nagoya University, Nagoya, Japan, in 1991, 1993, and 2000, respectively. In 1993, he joined SECOM IS Laboratory, Tokyo, Japan, and in 2000, Nara Institute of Science and Technology, Ikoma, Japan. Since 2014, he has been a Professor with The University of Tokyo, Tokyo, Japan. His research interests include statistical audio signal processing, blind source separation, and speech enhancement. He has put his research into the world's first commercially available independent-component-analysis-based BSS microphone in 2007. He was the recipient of several paper awards from IEICE in 2001 and 2006, from TAF in 2004, 2009, 2012, and 2018, from IEEE-IROS2005 in 2006, and from APSIPA in 2013 and 2018, and also the DOCOMO Mobile Science Award in 2011, Ichimura Award in 2013, Commendation for Science and Technology by the Minister of Education in 2015, Achievement Award from IEICE in 2017, and Hoko-Award in 2018. He has been professionally involved in various volunteer works for IEEE, EURASIP, IEICE, and ASJ. Since 2018, he has been an APSIPA Distinguished Lecturer.
\end{IEEEbiography}

\end{document}